\documentclass[aos]{imsart}

\usepackage{amsthm,amsmath, amssymb}
\usepackage{graphicx}
\usepackage{threeparttable}

\startlocaldefs

\newtheorem{propn}{Proposition}

\newtheorem{alg}{Algorithm}

\endlocaldefs

\begin{document}

\begin{frontmatter}

\title{Parallel marginalization Monte Carlo with applications to conditional path sampling}

\author{\fnms{Jonathan} \snm{weare}\corref{}\ead[label=e1]{weare@cims.nyu.edu}}
\address{Courant Institute, NYU\\ 251 Mercer St.\\ New York, NY 10012 \\ \printead{e1}}
 \affiliation{Department of Mathematics, University of California and Lawrence Berkeley National
 Laboratory, Berkeley, CA 94720 }


\runauthor{J. Weare}

\begin{abstract}
Monte Carlo sampling methods often suffer from long correlation times.
Consequently, these methods must be run
for many steps to generate an independent sample.  In this paper a method is proposed to overcome this difficulty.
The method utilizes information from rapidly equilibrating coarse Markov chains 
that sample marginal distributions of the full system.  This is accomplished through
 exchanges between the full chain and the auxiliary coarse chains.
Results of numerical tests on 
the bridge sampling and filtering/smoothing problems for a stochastic differential
equation are presented.
\end{abstract}



\end{frontmatter}

\section{Introduction}
In spite of substantial effort to improve the efficiency of Markov chain Monte Carlo (MCMC) methods, spatial correlations remain a major impediment.  These correlations can severely restrict the possible configurations of a system by imposing  complicated relationships between variables. 
It is well known that judicious elimination of variables by 
renormalization can reduce long range correlations (see \cite{kadanoff00,binney92}).  
The remaining variables are distributed according to
the marginal distribution,
$$
\overline{\pi}\left(x\right) = \int \pi\left(x,y\right)dy,
$$
where $\pi\left(x,y\right)$ is the full distribution.  Given the values of the $x$ variables and the
marginal distribution $\overline{\pi}$ the $y$ variables are distributed according to the conditional distribution
$$
\pi\left(y\vert x\right) = \frac{\pi\left(x,y\right)}{\overline{\pi}\left(x\right)}.
$$
For systems exhibiting critical phenomena, the 
path through the space of distributions taken 
by marginal distributions under repeated renormalization can yield 
essential information about critical indices and the location of 
critical points (see \cite{kadanoff00,binney92}).  
More generally, because these marginal distributions exhibit shorter correlation
lengths and weaker local correlations, 
they are useful in the acceleration of Markov
chain Monte Carlo methods.
As explained in the next section, parallel marginalization takes advantage
of the shorter correlation lengths present in
 marginal distributions of the target density.

The use of Monte Carlo updates on lower dimensional spaces  is not
a new concept.  In fact this is a necessary procedure in high dimensions.  One simply constructs a chain with steps that preserve the conditional probability density of the full measure.  This is usually accomplished
by perturbing a few components of the chain while holding all other components of the chain constant.
In other words the chain takes steps of the  form
$$
Y^{n+1} = (x_1,\dots,x_{i-1},x_i+\epsilon,x_{i+1},\dots,x_d)
$$
where 
$$
Y^n = (x_1,\dots,x_d)
$$
and the move preserves $\pi(x_i\vert x_1,\dots,x_{i-1},x_{i+1},\dots,x_d).$
There have been many important attempts to use proposals in more general sets of projected coordinates.
The multi-grid Monte Carlo method presented in \cite{goodman86,goodman89}
is one such method.
These techniques do not incorporate marginal densities.

In \cite{brandt01}, Brandt and Ron propose a multi-grid method which approximates 
successive marginal distributions of the Ising model and then uses 
these approximations to generate large scale movements of the Markov chain 
sampling the full joint distribution of all variables.  Their method, while
demonstrating the efficacy of incorporating information from successive
marginal distributions, suffers from two limitations.
First, the method used to approximate the marginal distributions is specific
to a small class of problems.  For example, it cannot be easily generalized 
to systems in continuous spaces.  Second, information from the 
approximate marginal distributions is adopted by the Markov chain 
in a way which does not preserve the target distribution of all variables.

The design of a generally applicable method which approximates the marginal 
distributions was addressed in \cite{chorin03, chorin06b} by Chorin, and in \cite{stinis05}
 by Stinis.
Both authors approximate the renormalized Hamiltonian of the system
given by the formula,
$$
\overline{\mathcal{H}}\left(x\right) =
-\log\int \pi\left(x,y\right)dy.
$$
Thus $\exp\left(-\overline{\mathcal{H}}\left(x\right)\right)$ 
is the marginal 
distribution of the $x$ variables.  Chorin determines the 
coefficients in an expansion of $\overline{\mathcal{H}}\left(x\right)$ by first
expanding the derivatives 
$\frac{\partial \overline{\mathcal{H}}\left(x\right)}{\partial x}$,
which can be expressed as conditional expectations with respect to the
full distribution.  Stinis shows that a maximum likelihood approximation
to the renormalized Hamiltonian can be found by minimizing the error
in the expectations of the basis functions in an expansion of 
$\overline{\mathcal{H}}\left(x\right)$.  For applications of related ideas to MCMC simulations
see \cite{okunev05} and \cite{chorin07}.

Two Parallel marginalization algorithms are developed in the next section along with propositions that guarantee  that the resulting Markov chains satisfy the detailed balance condition.
In the final section
the conditional path sampling problem is described and numerical results
are presented for the bridge sampling and smoothing/filtering problems.  A brief
introduction to parallel marginalization can be found in \cite{weare07}.

\section{Parallel marginalization}
In this section, it is assumed that appropriate 
approximate marginal distributions are available.
How to find these marginal distributions depends on the application and
will be discussed here only in the context of the examples presented in this paper.  
A new Markov chain Monte
Carlo method is introduced which uses approximate 
marginal distributions of the target distribution
to accelerate sampling.
Auxiliary 
Markov chains that sample approximate marginal distributions are evolved
simultaneously with the Markov chain that samples the distribution of interest.
By swapping their configurations, these auxiliary 
chains pass information between themselves and with the chain sampling
the original
distribution.

Assume that
the system of interest has
a probability density, $\pi_0(x_0)$, where
$x_0$ lies in some space $E.$
Suppose further that, by the Metropolis-Hastings or any other method (see \cite{liu02}),
one can construct a Markov chain, $Y_0^n\in E$, which has
$\pi_0$ as its stationary measure.  That is, for two points $x_0,y_0\in E$
$$
\int \tau_0(y_0 \vert x_0)\pi_0(x_0)\ dx_0 = \pi_0(y_0)
$$
where $\tau_0(y_0\vert x_0)$ is the probability density of a move
to $\left\{Y_0^{n+1}=y_0\right\}$ given that $\left\{Y_0^n=x_0\right\}$.  
Here, $n$ is the algorithmic step. 

In order to take advantage of the shorter spatial correlations exhibited by marginal distributions of 
$\pi_0$, a collection of lower dimensional Markov chains which approximately 
sample marginal distributions of $\pi_0$ is considered.
Suppose the random variable $X_0$ has $d_0$ components.  Divide
these into two subsets,
$$
X_0 = \left(\widehat{X}_0,\widetilde{X}_0\right),
$$
where $\widehat{X}_0$ has $d_1$ components and $\widetilde{X}_0$ has $d_0-d_1$
components.
Recall that the $\widehat{X}_0$ variables are distributed according to the
marginal density,
\begin{equation}\label{marginal}
\overline{\pi}_0(\hat{x}_0) = \int \pi_0(\hat{x}_0,\tilde{x}_0)d\tilde{x}_0
\end{equation}
and that given the value of the $\widehat{X}_0$ variables, the
$\widetilde{X}_0$ variables are distributed according to the conditional density,
\begin{equation}\label{conditional}
\pi(\tilde{x}_0\vert \hat{x}_0) = \frac{\pi_0(\hat{x}_0,\tilde{x}_0)}{\overline{\pi}_0(\hat{x}_0)}
\end{equation}
Label the domain of the $\widehat{X}_0$ variables $E_1$.
Suppose further that an approximation to the marginal distribution of the 
$\widehat{X}_0$ variables,
$$
\pi_1\left(\hat{x}_0\right)\approx \overline{\pi}_0\left(\hat{x}_0\right)
$$
is available.   The sense in which $\pi_1$ approximates $\overline{\pi}_0$ is intentionally left vague.
In applications of parallel marginalization the accuracy of the approximation manifests itself through an acceptance rate.

Now let $X_1\in E_1$ be independent of the $X_0$ random variables and
drawn from $\pi_1\left(\widehat{x}_0\right)$. 
 Notice that $X_1$ represents the same physical variables as
$\widehat{X}_0$ though its probability density is not the exact marginal density.
Continue in this way to remove variables from the system by decomposing
$X_l\in E_l$ into proper subsets as
$$
X_l=\left( \widehat{X}_l, \widetilde{X}_l \right)
$$
and defining $X_{l+1}\in E_{l+1}$ to be independent of the $\left\{X_0,\dots,X_l\right\}$ random variables
and drawn from an approximation $\pi_{l+1}$ to $\overline{\pi}_l\left(\hat{x}_l\right)$. 
Clearly each $X_{l+1}$ represents fewer physical variables than $X_l$.

Just as one can construct a Markov chain $Y^n_0\in E_0$ to sample $X_0$, one can also
 construct Markov chains
$Y_l^n\in E_l$ to sample $\pi_l$.
In other words, for each $Y_l^n$
 choose a transition probability density $\tau_l,$
such that 
$$
\int \tau_l(y_l\vert x_l) \pi_l\left(x_l\right) dx_l = \pi_l\left(y_l\right)
$$
for all $i.$

The chains $Y^n_l$ can be arranged in parallel to yield a larger Markov chain,
$$Y^n = \left(Y^n_0,\dots, Y^n_L\right)
\in  E_0
\times\dots\times E_L.
$$
The probability density of a move
to $\left\{Y^{n+1}=y\right\}$ given that $\left\{Y^n=x\right\}$
for $x,y\in E_0
\times\dots\times E_L$ is given by
\begin{equation}\label{def:tau}
\tau(y\vert x) = \prod_{l=0}^L \tau_l(y_l\vert x_l).
\end{equation}
Since
\begin{equation*}
\int \left(\tau(y\vert x) \prod_{l=0}^L\pi_l\left(x_l\right)\right)
\ dx_0\dots dx_L
= 
\prod_{l=0}^L\pi_l\left(y_l\right)
\end{equation*}
the stationary distribution of $Y^n$ is
\begin{equation*}
\Pi\left(x_0,\dots,x_L\right) = \pi_0\left(x_0\right)
\dots\pi_{L}\left(x_{L}\right).
\end{equation*}

The next step in the construction is to allow interactions between the chains
$Y_l^n$ and to thereby pass information from the rapidly 
equilibrating chains on the lower dimensional spaces (large $l$) down
to the chain on the original space ($l=0$).
This is accomplished by swap moves.
In a swap move between levels $l$ and $l+1$, a subset, 
$\hat{x}_l\in E_{l+1}$, of the $x_l$ variables is exchanged with
the $x_{l+1}\in E_{l+1}$ variables.  The remaining $\tilde{x}_l$ variables are resampled
from the conditional distribution $\pi_l\left(\tilde{x}_l\vert x_{l+1}\right)$.
For the full chain, this swap takes the form of a move from $\left\{Y^n=x\right\}$ 
to $\left\{Y^{n+1}=y\right\}$ where
$$
x = \left(\dots,\hat{x}_l,\tilde{x}_l,x_{l+1},\dots\right)
$$
and
$$
y = \left(\dots,x_{l+1},\tilde{y}_l,\hat{x}_l,\dots\right).
$$
The $\tilde{y}_l$ variables are drawn from $\pi_l\left(\tilde{x}_l\vert x_{l+1}\right)$ and
the ellipses represent components of $Y^n$ that remain unchanged in the 
transition.

If these swaps are undertaken unconditionally, the resulting chain may
equilibrate rapidly, but will not, in general, preserve the product distribution
$\Pi$.  To remedy this  the swap acceptance probability 
\begin{equation}\label{def:A_l}
A_l =
\min\biggl\lbrace 1,\ \frac{ \overline{\pi}_l(x_{l+1})\pi_{l+1}(\hat{x}_l)}
{\overline{\pi}_l(\hat{x}_l)\pi_{l+1}(x_{l+1})}\biggr\rbrace
\end{equation}
is introduced.
Recall that  $\overline{\pi}_l$ is the function 
resulting from the integration of $\pi_l$ over the $\tilde{x}_l$ variables 
as in equation \eqref{marginal}.
Given that $\left\{Y^n=x\right\}$, the probability density of 
$\left\{Y^{n+1}=y\right\}$, after the proposal and
either acceptance with probability $A_l$ or rejection
with probability $1-A_l$, of a swap move, is given by
\begin{multline*}
\psi_l\left(y\vert x\right) =
 \left(1-A_l\right)\ 
\prod \delta_{\left\{y_j=x_j\right\}}\\
+ A_l\ \pi_l(\tilde{y}_l \vert x_{l+1})\ 
\delta_{\left\{\left(\hat{y}_l,y_{l+1}\right)
=\left(x_{l+1},\hat{x}_l\right)\right\}}
\prod_{j\notin\left\{l,l+1\right\}}\delta_{\left\{y_j=x_j\right\}}
\end{multline*}
for $x,y \in E_0
\times\dots\times E_L$. $\delta$ is the Dirac delta function.

We have the following proposition.
\begin{propn}
The transition probabilities $\psi_l$ satisfy the
detailed balance condition for the measure $\Pi,$ 
i.e.
\begin{equation*}
\Pi(x)\ \psi_l\left(y\vert x\right) = 
\Pi(y)\ \psi_l\left(x\vert y\right)
\end{equation*}
where $x,y \in E_0
\times\dots\times E_L.$
\end{propn}
\begin{proof}
Fix $x,y \in E_0
\times\dots\times E_L$ such that $x\neq y$.
\begin{multline*}
\Pi(x)\ \psi_l\left(y\vert x\right)  = 
\left(\prod_{j\notin\left\{l,l+1\right\}}  \pi_j\left(x_j\right)\delta_{\left\{y_j=x_j\right\}}  \right)
\pi_l\left(x_l\right)\pi_{l+1}\left(x_{l+1}\right)\\
\times \left(\left(1-A_l\right)\ 
 \delta_{\left\{\left(y_l,y_{l+1}\right)=\left(x_l,x_{l+1}\right)\right\}}
 + A_l\ \pi_l(\tilde{y}_l \vert x_{l+1})\ 
\delta_{\left\{\left(\hat{y}_l,y_{l+1}\right)
=\left(x_{l+1},\hat{x}_l\right)\right\}}\right)
\end{multline*}

When $x\neq y$ $\left(\Pi(x)\psi_l\left(y \vert x\right)\right)$ and  $\left(\Pi(y)\psi_l\left(x\vert y\right)\right)$ are
both zero unless 
$x_j=y_j$ for all $j$ except $l$ and $l+1$ and 
$\left(\hat{y}_l,y_{l+1}\right)=\left(x_{l+1},\hat{x}_l\right)$.
Therefore it is enough to check that the function
$$
R\left(\left(x_l,x_{l+1}\right),\left(y_l,y_{l+1}\right)\right) = \pi_l\left(x_l\right)\pi_{l+1}\left(x_{l+1}\right)
 \pi_l\left(\tilde{y}_l \vert x_{l+1}\right)\ A_l
$$
is symmetric in $\left(x_l,x_{l+1}\right)$ and $\left(y_l,y_{l+1}\right)$
when $\left(\hat{y}_l,y_{l+1}\right)=\left(x_{l+1},\hat{x}_l\right)$.
Plugging in the definition of $A_l,$
\begin{equation*}
R
 =
  \pi_l\left(x_l\right)\pi_{l+1}\left(x_{l+1}\right)\pi_l\left(\tilde{y}_l \vert x_{l+1}\right)\ 
\min\biggl\lbrace 1,\ \frac{ \overline{\pi}_l(x_{l+1})\pi_{l+1}(\hat{x}_l)}
{\overline{\pi}_l(\hat{x}_l)\pi_{l+1}(x_{l+1})}\biggr\rbrace
\end{equation*}
Rearranging terms gives,
\begin{align*}
R &=  \pi_l\left(x_l\right)\pi_{l+1}\left(x_{l+1}\right)
\pi_l\left(\tilde{y}_l \vert x_{l+1}\right)\overline{\pi}_l\left(x_{l+1}\right)\pi_{l+1}\left(y_{l+1}\right)\\
& \hspace{24pt} \times
\min\biggl\lbrace \frac{1}{ \overline{\pi}_l(x_{l+1})\pi_{l+1}(y_{l+1})},\ \frac{1}
{\overline{\pi}_l(y_{l+1})\pi_{l+1}(x_{l+1})}\biggr\rbrace
\end{align*}
Recall from \eqref{conditional}, 
that $\pi_l(\tilde{y}_l\vert x_{l+1})\overline{\pi}_l(x_{l+1}) = \pi_l(x_{l+1},\tilde{y}_l).$
Therefore, since $x_{l+1}=\hat{y}_l,$
\begin{align*}
R &=  \pi_l\left(x_l\right)\pi_{l+1}\left(x_{l+1}\right)
\pi_l\left(y_l\right) \pi_{l+1}\left(y_{l+1}\right)\\
& \hspace{24pt} \times
\min\biggl\lbrace \frac{1}{ \overline{\pi}_l(x_{l+1})\pi_{l+1}(y_{l+1})},\ \frac{1}
{\overline{\pi}_l(y_{l+1})\pi_{l+1}(x_{l+1})}\biggr\rbrace\\
\end{align*}
The final formula is clearly symmetric in $\left(x_l,x_{l+1}\right)$ and $\left(y_l,y_{l+1}\right)$.
\end{proof}
The detailed balance condition stipulates that the probability of
observing a transition $x\rightarrow y$ is equal to that
of observing a transition $y\rightarrow x$ and guarantees that
the resulting Markov Chain preserves the distribution $\Pi$.
If the chain is also Harris recurrent then 
 averages over a trajectory of $Y^n$ 
will converge to averages over $\Pi$.  
In fact,  chains generated  by swaps as described above cannot
be recurrent and must be combined with another transition rule to generate a convergent
Markov chain.
Since
$$
\pi_0(x_0) = \int \Pi(x_0,\dots,x_L)\ dx_1\dots dx_L,
$$
if $Y^n$ is Harris recurrent with invariant distribution $\Pi$, 
averages over $\pi_0$ can be calculated by taking averages over the trajectories 
of the first $d_0$ components of $Y^n$.

\subsection{Approximation of acceptance probabilities}
Notice that the formula \eqref{def:A_l} for $A_l$ requires the evaluation of
$\overline{\pi}_l$ at the points $\hat{x}_l,x_{l+1}\in E_{l+1}.$
While the approximation of 
$\overline{\pi}_l$ by functions on $E_{l+1}$
is in general a very difficult 
problem, its evaluation at a single point is often not 
terribly demanding.  In fact, in many cases, including the examples in
Chapter 3, the $\widehat{X}_l$ variables can be chosen so that the remaining
$\widetilde{X}_l$ variables are conditionally independent given 
$\widehat{X}_l.$

Despite this mitigating factor, the requirement that
$\overline{\pi}_l$ be evaluated before acceptance of any swap
is inconvenient.  Fortunately, and somewhat surprisingly, 
this requirement is not necessary.
In fact, standard strategies for approximating the
point values of the marginals yield Markov chains that also preserve 
the target measure.  Thus even a poor estimate of the ratio appearing
in \eqref{def:A_l} can give rise to a method that is exact in the sense that the
resulting Markov chain will asymptotically sample the target measure.

Before moving on to the description of the resulting Markov chain Monte Carlo algorithms 
consider briefly the general problem of evaluating  marginal densities.
Let $p_1(x,y)$ and $p_2(x,y)$ be the densities of two equivalent measures with 
marginal densities,
$$
\overline{p}_1(x) = \int p_1(x,y) dy
$$
and
$$
\overline{p}_2(x) = \int p_2(x,y) dy
$$
respectively.
For any integrable function $\gamma(x,y),$ 
\begin{align*}
\mathbf{E}_{p_1}\left[ \gamma\left(X,Y\right) p_2\left(X,Y\right) 
	\vert \left\{X=x\right\}\right]
&= \int \gamma(x,y) p_2(x,y) p_1(y\vert x) dy\\
&= \frac{\overline{p}_2(x)}{\overline{p}_1(x)} \int \gamma(x,y) p_2(y\vert x) p_1(x,y) dy\\
&= \frac{\overline{p}_2(x)}{\overline{p}_1(x)}
\mathbf{E}_{p_2}\left[ \gamma\left(X,Y\right) p_1\left(X,Y\right) 
	\vert \left\{X=x\right\}\right]
\end{align*}

Thus  given 
$\overline{p}_2(x),$  the value of $\overline{p}_1$ at $x$ can be obtained through 
the formula,
\begin{equation}\label{margformula}
\overline{p}_1(x) = \overline{p}_2(x)
\frac{\mathbf{E}_{p_2}\left[ \gamma\left(X,Y\right) p_1\left(X,Y\right) 
	\vert \left\{X=x\right\}\right]}
	{\mathbf{E}_{p_1}\left[ \gamma\left(X,Y\right) p_2\left(X,Y\right) 
	\vert \left\{X=x\right\}\right]}
\end{equation}
Of course, the usual importance sampling concerns  
apply here.  In particular, the approximation of the conditional expectations in \eqref{margformula}
 will be much easier when $Y$ lives in a lower dimensional space.

Similar approximations can be inserted into our acceptance probabilities $A_l$
in place of the ratio $
 \frac{ \overline{\pi}_l(x_{l+1})}
{\overline{\pi}_l(\hat{x}_l)}.
$
For example, if $p_l(\tilde{x}_l \vert \hat{x}_l)$ is a reference density approximating
$\pi_l(\tilde{x}_l \vert \hat{x}_l),$ then the choice
$$
\gamma(\hat{x}_l,\tilde{x}_l)=\frac{1}{p_l(\hat{x}_l,\tilde{x}_l)}
$$
yields 
\begin{equation}\label{margapprox}
\overline{\pi}_l(\hat{x}) \approx \overline{p}_l(\hat{x}) 
\frac{1}{M} \sum \frac{\pi_l\left(\hat{x}_l,V^j\right)}{p_l\left(\hat{x},V^j\right)}
=
 \frac{1}{M} \sum \frac{\pi_l\left(\hat{x}_l,V^j\right)}{p_l\left(V^j\vert\hat{x}_l\right)}
\end{equation}
where the $V^j$ are samples from $p_l(\tilde{x}_l\vert \hat{x}_l).$
Thus if $U^j$ are samples from $p_l(\tilde{x}_l \vert x_{l+1}),$ then
\begin{equation*}
\frac{\frac{1}{M}\sum_{j=1}^{M} \frac{\pi_l\left(x_{l+1},U^j\right)}
{p_l\left(U^j\vert x_{l+1}\right)}}
{\frac{1}{M}\sum_{j=1}^{M} \frac{\pi_l\left(\hat{x}_l,V^j\right)}
{p_l\left(V^j\vert \hat{x}_l\right)}}
\xrightarrow[{M\rightarrow\infty}]{a.s.}
\frac{\mathbf{E}_{p_l}\left[\frac{\pi_l\left(x_{l+1},\widetilde{X}_l\right)}
{p_l\left(\widetilde{X}_l\vert x_{l+1}\right)}\ \vert
\left\{\widehat{X}_l=x_{l+1}\right\}\right]}
{\mathbf{E}_{p_l}\left[\frac{\pi_l\left(\hat{x}_l,\widetilde{X}_l\right)}
{p_l\left(\widetilde{X}_l\vert \hat{x}_l\right)}\ \vert
\left\{\widehat{X}_l=\hat{x}_l\right\}\right]}
= \frac{\overline{\pi}_l(x_{l+1})}{\overline{\pi}_l(\hat{x}_l)}
\end{equation*}
  In the numerical examples presented here, $p_l(\ \cdot\ \vert\hat{x}_l)$
  is a Gaussian approximation of 
  $\pi_l(\tilde{x}_l\vert \hat{x}_l)$.
  How $p_l$ is chosen depends on the problem at hand 
  (see numerical examples below).  
  In general $p_l(\ \cdot\ \vert\hat{x}_l)$ should be easily evaluated and 
  independently sampled, and it should ``cover'' $\pi_l(\ \cdot\ \vert \hat{x}_l)$
  in the sense that regions
  where $\pi_l(\ \cdot\ \vert \hat{x}_l)$
  is not negligible
  should be contained in regions where $p_l(\ \cdot\ \vert\hat{x}_l)$ is not negligible. 
  In the case mentioned above that the $\widehat{X}_l$ variables can be chosen so that
  the remaining $\widetilde{X}_l$ variables are conditionally independent given 
  $\widehat{X}_l$ the conditional density $\pi_l(\tilde{x}_l\vert \hat{x}_l)$
   can be written as a product of many
  low dimensional densities.
  As mentioned above, the problem of finding a reference density for
  importance sampling is much simpler in low dimensional spaces.

The following algorithm results from replacing $A_l$ in \eqref{def:A_l} with  approximation of the form \eqref{margapprox}.
Assume that the current position
of the chain is $\left\{Y^n=x\right\}$ where
$$
x = \left(\dots,\hat{x}_l,\tilde{x}_l,x_{l+1},\dots\right).
$$
Algorithm \ref{pm1} will result in either $\left\{Y^{n+1}=x\right\}$
or $\left\{Y^{n+1}=y\right\}$ where
$$
y = \left(\dots,x_{l+1},\tilde{y}_l,\hat{x}_l,\dots\right)
$$
and $\tilde{y}_l$ is approximately drawn from $\pi_l\left(\tilde{x}_l\vert x_{l+1}\right).$

\begin{alg}[Parallel Marginalization 1]\label{pm1}
The chain moves from $Y^n$ to $Y^{n+1}$ as follows:
\begin{enumerate}
\item
Let $U^j$ for 
$j=1,\dots,M$ be independent
random variables sampled from $p_l(\ \cdot\ \vert x_{l+1})$  (recall
that the swap is between $\hat{x}_l$ and ${x_{l+1}}$ which
are both in $E_{l+1}$).
\item
Evaluate the weights
$$
W_U^j = \frac{\pi_l\left(x_{l+1},U^j\right)}
{p_l\left(U^j\vert\ x_{l+1}\right)}.
$$
The choice of $p_l$ made above affects the variance of these weights, and therefore
the variance of the acceptance probability below.
\item Draw the random index $J\in\left\{1,\dots,M\right\}$  according to the
probabilities
$$
\mathbf{P}\left[ J = j \right] 
= \frac{W_U^j}{\sum_{m=1}^{M} W_U^m}.
$$
Set $\widetilde{Y}'=U^J.$
Notice that  $\widetilde{Y}'$ is an approximate sample from $\pi_l(\ \cdot\ \vert x_{l+1}).$ 
\item Let $V^J=\tilde{x}_l$  
  and draw $V^j$ 
  for $j\neq J$  independently
from $p_l(\ \cdot\ \vert\hat{x}_l),$
Notice that the
$U^j$ variables depend on $x_{l+1}$
 while the $V^j$ variables depend on $\hat{x}_l$.
\item Define the weights
$$
W_V^j = \frac{\pi_l\left(\hat{x}_l,V^j\right)}
{p_l\left(V^j\vert\hat{x}_l\right)}
$$
\item Set
$$Y^{n+1} = \left(\dots,x_{l+1},\widetilde{Y}',\hat{x}_l,\dots\right)$$
with probability
\begin{equation}\label{def:A^M_l}
A^M_l = \min\biggl\lbrace 1,\ \frac{ \pi_{l+1}(\hat{x}_l)\sum_{j=1}^{M} 
W^j_U}
{\pi_{l+1}(x_{l+1})
\sum_{j=1}^{M} W^j_V}\biggr\rbrace
\end{equation}
and
$$Y^{n+1} = Y^n=\left(\dots,\hat{x}_l,\tilde{x}_l,x_{l+1},\dots\right)$$
with probability
$
1-A^M_l
$.
\end{enumerate}
\end{alg}
The transition probability density for the above swap move from
$x$ to  $y$ for $x,y \in E_0
\times\dots\times E_L$is given by
\begin{multline*}
\psi^M_l(y\vert x) = 
\mathbf{P}\left[\left\{\text{Swap is rejected}\right\}\right]\ 
\prod \delta_{\left\{y_j=x_j\right\}}\\
+\mathbf{P}\left[\left\{\text{Swap is accepted}\right\}\cap 
\left\{\widetilde{Y}'=\tilde{y}_l\right\} \right]\\
\times \delta_{\left\{\left(\hat{y}_l,y_{l+1}\right)
=\left(x_{l+1},\hat{x}_l\right)\right\}}
\prod_{j\notin\left\{i,i+1\right\}}\delta_{\left\{y_j=x_j\right\}},
\end{multline*}
where
 $\delta$ is again the Dirac delta function.
Notice that to find the probability density
$\mathbf{P}\left[\left\{\text{Swap is accepted}\right\}\cap 
\left\{\widetilde{Y}'=\tilde{y}_l\right\} \right]$ 
 one must integrate over
the possible values of the $U^j$ and $V^j$ variables.  
Since $\pi_l$ appears in the integrand
it is not possible, in general, to evaluate the integral.   However, as indicated in the
proof of the next proposition, it is not necessary to evaluate this density to show that the
method converges.

While the preceding swap move corresponds to a
method for approximating the ratio
$$
 \frac{ \overline{\pi}_l(x_{l+1})}
{\overline{\pi}_l(\hat{x}_l)}
$$
appearing in formula \eqref{def:A_l} for $A_l,$ it also
has similarities with
the multiple-try Metropolis method,
presented in \cite{liu00,qin01}, that uses multiple suggestion samples
to improve acceptance rates of standard MCMC methods.
In fact the proof of the following proposition is motivated by the  proof of the detailed balance
condition for the multiple try method.
\begin{propn}
The transition probabilities $\psi^M_l$ satisfy the
detailed balance condition for the measure $\Pi.$ 
\end{propn}
\begin{proof}
For $x,y \in E_0
\times\dots\times E_L$ such that $x\neq y$,
\begin{multline*}
\Pi(x)\ \psi^M_l(y\vert x) = 
\Pi(x)\  \mathbf{P}\left[\left\{\text{Swap is accepted}\right\}\cap 
\left\{\widetilde{Y}'=\tilde{y}_l\right\} \right]\\
\times \delta_{\left\{\left(\hat{y}_l,y_{l+1}\right)
=\left(x_{l+1},\hat{x}_l\right)\right\}}
\prod_{j\notin\left\{i,i+1\right\}}\delta_{\left\{y_j=x_j\right\}},
\end{multline*}
As in the previous proof it can be assumed that $x_j=y_j$ for all $j$ except $l$ and $l+1$ and 
$\left(\hat{y}_l,y_{l+1}\right)=\left(x_{l+1},\hat{x}_l\right)$.
Since in this case $\pi(x_j)=\pi(y_j)$ for all $j\notin\left\{l,l+1\right\},$ 
it remains to show that if $\left(\hat{y}_l,y_{l+1}\right)=\left(x_{l+1},\hat{x}_l\right)$
then
\begin{multline*}
R\left(\left(x_l,x_{l+1}\right),\left(y_l,y_{l+1}\right)\right) 
= \pi_l\left(x_l\right)\pi_{l+1}\left(x_{l+1}\right)\\
\times\mathbf{P}\left[\left\{\text{Swap is accepted}\right\}\cap 
\left\{\widetilde{Y}^{n+1}_l=\tilde{y}_l\right\} \right]
\end{multline*}
is symmetric in $\left(x_l,x_{l+1}\right)$ and $\left(y_l,y_{l+1}\right)$.
Define a random index  $J\in\left\{1,\dots,M\right\}$ 
by the relation
$\tilde{y}_l=U^J.$
Then, since the $U^j$ are $i.i.d.$,
\begin{align*}
\mathbf{P}\left[\left\{\text{Swap is accepted}\right\}\cap 
\left\{\widetilde{Y}'=\tilde{y}_l\right\} \right]
= & \\
& \hspace{-70pt} \sum_{j=1}^{M}\mathbf{P}\left[\left\{\text{Swap is accepted}\right\}\cap
 \left\{\tilde{y}_l=U^j \right\} \cap \left\{J=j\right\} \right]\\
 & \hspace{-70pt} = M\ \mathbf{P}\left[\left\{\text{Swap is accepted}\right\}\cap
 \left\{\tilde{y}_l=U^1 \right\} \cap \left\{J=1\right\} \right]
\end{align*}
Thus,
\begin{align*}
R\left(\left(x_l,x_{l+1}\right),\left(y_l,y_{l+1}\right)\right)  = &\ 
M\ \pi_l\left(x_l\right)\pi_{l+1}\left(x_{l+1}\right)\\
&\times\mathbf{P}\left[\left\{\text{Swap is accepted}\right\}\cap
 \left\{\tilde{y}_l=U^1 \right\} \cap \left\{J=1\right\} \right]
 \end{align*}
  Writing out the density on the right of this relation gives,
 \begin{align*}
 R =&\ M\ \pi_l\left(x_l\right)\pi_{l+1}\left(x_{l+1}\right)
 \int
 \min\biggl\lbrace 1,\ \frac{ \pi_{l+1}(\hat{x}_l)\sum_{j=1}^{M} 
W^j_U}
{\pi_{l+1}(x_{l+1})
\sum_{j=1}^{M} W^j_V}\biggr\rbrace
\frac{\frac{\pi_l\left(x_{l+1},u^1\right)}{p\left(u^1\vert x_{l+1}\right)}}{\sum_{j=1}^{M}W^j_U}\\
&\hspace{24pt}\times p\left(u^1\vert x_{l+1}\right)
\prod_{j>1} p\left(u^j\vert x_{l+1}\right)p\left(v^j\vert \hat{x}_l\right)du^j dv^j
\end{align*}
Replacing $u^1$ by $\tilde{y}_l$ and rearranging gives,
\begin{align*}
R 
=&\ M\ \pi_l\left(x_l\right)\pi_{l+1}\left(x_{l+1}\right)
\pi_l\left(x_{l+1},\tilde{y}_l\right)\pi_{l+1}\left(\hat{x}_l\right)\\
&\hspace{24pt}\times \int
 \min\biggl\lbrace \frac{1}{\pi_{l+1}\left(\hat{x}_l\right)
 \sum_{j=1}^{M}W^j_U},\ \frac{ 1}
{\pi_{l+1}(x_{l+1})
\sum_{j=1}^{M} W^j_V}\biggr\rbrace\\
&\hspace{48pt}\times 
\prod_{j>1} p\left(u^j\vert x_{l+1}\right)p\left(v^j\vert \hat{x}_l\right)du^j dv^j.
\end{align*}
Since $x_{l+1}=\hat{y}_l,$ $\pi_l(x_{l+1},\tilde{y}_l)=\pi_l(y_l).$
Therefore, after replacing $\hat{x}_l$ by $y_{l+1},$
\begin{align*}
R 
=&\ M\ \pi_l\left(x_l\right)\pi_{l+1}\left(x_{l+1}\right)
\pi_l\left(y_l\right)\pi_{l+1}\left(y_{l+1}\right)\\
&\hspace{24pt}\times \int
 \min\biggl\lbrace \frac{1}{\pi_{l+1}\left(y_{l+1}\right)
 \sum_{j=1}^{M}W^j_U},\ \frac{ 1}
{\pi_{l+1}(x_{l+1})
\sum_{j=1}^{M} W^j_V}\biggr\rbrace\\
&\hspace{48pt}\times 
\prod_{j>1} p\left(u^j\vert x_{l+1}\right)p\left(v^j\vert y_{l+1}\right)du^j dv^j.
\end{align*}
which is symmetric in $\left(x_l,x_{l+1}\right)$ and $\left(y_l,y_{l+1}\right)$.
\end{proof}

For small values of $M$ in \eqref{def:A^M_l}, calculation of the swap acceptance probabilities
is very cheap. However, higher values of $M$ 
may improve the acceptance rates.  For example, if the $\pi_l$ for $l>0$ are
exact marginals of $\pi_0,$ then $A_l\equiv 1$ while $A^M_l\leq 1.$
In practice one has to balance
the speed of evaluating  $A_l^M$ for small $M$ with the possible higher acceptance rates
for $M$ large.

In analogy again with the multiple-try method, the above algorithm can be generalized to
include correlated samples $U^j$ and $V^j$.  This generalization is useful because it allows
reference densities that cannot be independently sampled.
Again consider a transition from 
$\left\{Y^n=x\right\}$ where
$$
x = \left(\dots,\hat{x}_l,\tilde{x}_l,x_{l+1},\dots\right)
$$
to
either $\left\{Y^{n+1}=x\right\}$
or $\left\{Y^{n+1}=y\right\}$ where
$$
y = \left(\dots,x_{l+1},\tilde{y}_l,\hat{x}_l,\dots\right).
$$

First choose some reference transition densities 
$p_l^j\left(u^j\vert \left(u^0,\dots,u^{j-1}\right),\ \hat{x}_l\right)$ that sample a variable $U^j$
given the previous $j-1$ samples and the value of the $\hat{x}_l$ variables.
Let 
\begin{equation}\label{prodp}
p_l^j\left((u^{k+1},\dots,u^j)\vert (u^0,\dots,u^k),\ \hat{x}_l\right)
= \prod_{k<m\leq j} p_l^m\left(u^m\vert (u^0,\dots,u^{m-1}),\ \hat{x}_l\right).
\end{equation}
For example, one might choose the $p^j_l$ to be  Markov transition kernels associated
with some Markov chain Monte Carlo method with stationary measure 
$\pi_l(\tilde{x}_l\vert \hat{x}_l).$
Also let $\lambda^j\left((u^0,\dots,u^j),\ \hat{x}_l,x_{l+1}\right)$ be any function
satisfying the relation
$$
\lambda^j\left((u^0,\dots,u^j),\ \hat{x}_l,x_{l+1}\right) = 
\lambda^j\left((u^j,\dots,u^0),\ x_{l+1},\hat{x}_l\right)
$$

\begin{alg}[Parallel Marginalization 2]\label{pm2}
We move the chain from $Y^n$ to $Y^{n+1}$ as follows:
\begin{enumerate}
\item 
For $j=1,\dots,M$ sample $U^j$ from 
$p_l^j\left(\ \cdot\ \vert (\tilde{x}_l, U^1,\dots,U^{j-1}),\ x_{l+1}\right).$
Notice the conditioning on the value $\widehat{X}_l=x_{l+1}$.
\item Define the weights
\begin{align*}
W_U^j &=  \pi_l\left(U^j,x_{l+1}\right) 
p_l^j\left(\left(U^{j-1},\dots,U^1,\tilde{x}_l\right)\vert U^j,\ \hat{x}_l\right)\\
&\hspace{24pt} \times\lambda^j\left(\left(\tilde{x}_l,U^1,\dots,U^j\right),\  \hat{x}_l,x_{l+1}\right).
\end{align*}
Notice the reversal in the ordering of the $U^j$ and the conditioning on 
$\widehat{X}_l=\hat{x}_l$.

\item Choose the random index $J\in\left\{1,\dots,M\right\}$ according to the
probabilities
$$
\mathbf{P}\left[ J=j \right] 
= \frac{W_U^j}{\sum_{m=1}^{M} W_U^m}.
$$
Set $\widetilde{Y}' = U^J.$
\item Let $V^J = \tilde{x}_l$ and for $j=1,\dots,J-1$
let $V^j=U^{J-j}$. 
For $j=J+1,\dots,M$
 sample $V^j$ from 
$p_l^j\left(\ \cdot\ \vert (\widetilde{Y}',\dots,V^{j-1}),\ \hat{x}_l\right).$
Notice the conditioning on the value $\widehat{X}_l=\hat{x}_l$.

\item Define the weights
\begin{align*}
W_V^j &= \pi_l\left(V^j,\hat{x}_{l}\right) 
p_l^j\left((V^{j-1},\dots,V^1,\widetilde{Y}')\vert V^j,\ {x}_{l+1}\right)\\
&\hspace{24pt}\times\lambda^j\left((\widetilde{Y}',V^1,\dots,V^j),\  x_{l+1},\hat{x}_l\right).
\end{align*}

\item Set
$$Y^{n+1} = \left(\dots,x_{l+1},\widetilde{Y}',\hat{x}_l,\dots\right)$$
with probability
\begin{equation}\label{def:A^M_l}
A^M_l = \min\biggl\lbrace 1,\ \frac{ \pi_{l+1}(\hat{x}_l)\sum_{j=1}^{M} 
W^j_U}
{\pi_{l+1}(x_{l+1})
\sum_{j=1}^{M} W^j_V}\biggr\rbrace
\end{equation}
and
$$Y^{n+1} = Y^n=\left(\dots,\hat{x}_l,\tilde{x}_l,x_{l+1},\dots\right)$$
with probability
$
1-A^M_l
$.
\end{enumerate}
\end{alg}

The transition probability density for the above swap move from
$x$ to  $y$ for $x,y \in E_0
\times\dots\times E_L$is again given by
\begin{multline*}
\psi^M_l(y\vert x) = 
\mathbf{P}\left[\left\{\text{Swap is rejected}\right\}\right]\ 
\prod \delta_{\left\{y_j=x_j\right\}}\\
+\mathbf{P}\left[\left\{\text{Swap is accepted}\right\}\cap 
\left\{\widetilde{Y}'=\tilde{y}_l\right\} \right]\\
\times  \delta_{\left\{\left(\hat{y}_l,y_{l+1}\right)
=\left(x_{l+1},\hat{x}_l\right)\right\}}
\prod_{j\notin\left\{i,i+1\right\}}\delta_{\left\{y_j=x_j\right\}}
\end{multline*}
where
and $\delta$ is again the Dirac delta function.  Again, the density \linebreak
$\mathbf{P}\left[\left\{\text{Swap is accepted}\right\}\cap 
\left\{\widetilde{Y}'=\tilde{y}_l\right\} \right]$ cannot and 
need not be evaluated.

Algorithm \ref{pm1} can be derived from Algorithm \ref{pm2} by setting
$$
p_l^j\left(u^j\vert (u^0,\dots,u^{j-1}),\ \hat{x}_l\right) = p_l\left(u^j\vert \hat{x}_l\right)
$$
and 
$$
\lambda^j\left((u^0,\dots,u^j),\ \hat{x}_l,x_{l+1}\right)
=\frac{1}{p_l\left(u^j\vert \hat{x}_l\right)p_l\left(u^0\vert x_{l+1}\right)}.
$$
Notice also that if 
\begin{align*}
\lambda^j\left((u^0,\dots,u^j),\hat{x}_l,x_{l+1}\right)&\\
&\hspace{-24pt}=\frac{q^j\left((u^1,\dots,u^{j-1})\vert \hat{x}_l,x_{l+1}\right)}
{p^j_l\left((u^{j-1},\dots,u^0) \vert u^j,\ \hat{x}_l\right)
p^j_l\left((u^{1},\dots,u^j)\vert u^0,\ x_{l+1}\right)},
\end{align*}
where, for each $j,$ $q^j$ is a conditional density satisfying 
$$
q^j\left((u^1,\dots,u^{j-1})\vert \hat{x}_l,x_{l+1}\right)
= q^j\left((u^{j-1},\dots,u^{1})\vert x_{l+1}, \hat{x}_l\right)
$$
then
$$
\mathbf{E}_{p_l^j}\left[W_U^j\right] = 
\int \pi_l(u^j,x_{l+1}) q^j\left((u_1,\dots,u_{j-1})\vert \hat{x}_l,x_{l+1}\right)\prod_{i=1}^j du^i  = 
\overline{\pi}_l(x_{l+1}).
$$
Thus, if the $p_l^j$ generate a sequence that satisfies a Law of Large Numbers, then 
$\frac{1}{M}\sum W_U^j\rightarrow \overline{\pi}_l(x_{i+1}).$
The same holds for the $W_V^j$ so that 
$$
A^M_l\rightarrow \min\biggl\lbrace 1,\ \frac{ \overline{\pi}_l(x_{l+1})\pi_{l+1}(\hat{x}_l)}
{\overline{\pi}_l(\hat{x}_l)\pi_{l+1}(x_{l+1})}\biggr\rbrace = A_l.
$$
More general choices of $\lambda^j$ lead to $A^M_l$ which converge to correpondingly more general acceptance probabilities than $A_l.$

Of course, expression \eqref{margformula} points the way to even more general algorithms.  Algorithms \ref{pm1} and \ref{pm2} correspond to choices of $\gamma$ in \eqref{margformula} that make the conditional expectation on the bottom of \eqref{margformula} equal to one.  Other choices of $\gamma$ may improve the variance of the resulting weights.

\begin{propn}
The transition probabilities $\psi^M_l$ satisfy the
detailed balance condition for the measure $\Pi.$ 
\end{propn}
\begin{proof}
Fix $x,y \in E_0
\times\dots\times E_L$ such that $x\neq y$.
For $x,y \in E_0
\times\dots\times E_L$ such that $x\neq y$,
\begin{multline*}
\Pi(x)\ \psi^M_l(y\vert x) = 
\Pi(x)\  \mathbf{P}\left[\left\{\text{Swap is accepted}\right\}\cap 
\left\{\widetilde{Y}'=\tilde{y}_l\right\} \right]\\
\times \delta_{\left\{\left(\hat{y}_l,y_{l+1}\right)
=\left(x_{l+1},\hat{x}_l\right)\right\}}
\prod_{j\notin\left\{i,i+1\right\}}\delta_{\left\{y_j=x_j\right\}},
\end{multline*}
As in the previous two proofs it can be assumed that $x_j=y_j$ for all $j$ except $l$ and $l+1$ and 
$\left(\hat{y}_l,y_{l+1}\right)=\left(x_{l+1},\hat{x}_l\right)$.
Since in this case $\pi(x_j)=\pi(y_j)$ for all $j\notin\left\{l,l+1\right\},$
  it remains to show that if $\left(\hat{y}_l,y_{l+1}\right)=\left(x_{l+1},\hat{x}_l\right)$
then
\begin{multline*}
R\left(\left(x_l,x_{l+1}\right),\left(y_l,y_{l+1}\right)\right) = \pi_l\left(x_l\right)\pi_{l+1}\left(x_{l+1}\right)\\ 
\times\mathbf{P}\left[\left\{\text{Swap is accepted}\right\}\cap 
\left\{\widetilde{Y}'=\tilde{y}_l\right\} \right]
\end{multline*}
is symmetric in $\left(x_l,x_{l+1}\right)$ and $\left(y_l,y_{l+1}\right)$.
Summing over disjoint events,
\begin{multline*}
\mathbf{P}\left[\left\{\text{Swap is accepted}\right\}\cap 
\left\{\widetilde{Y}'=\tilde{y}_l\right\} \right]
= \\
\sum_{j=1}^{M}\mathbf{P}\left[\left\{\text{Swap is accepted}\right\}\cap
 \left\{\tilde{y}_l=U^j \right\} \cap \left\{J=j\right\} \right]
\end{multline*}
Thus $R$ will be symmetric if for each $j$ the function
\begin{multline*}
R_j\left((x_l,x_{l+1}),(y_l,y_{l+1})\right) =  \pi_l\left(x_l\right)\pi_{l+1}\left(x_{l+1}\right)\\ 
\times\mathbf{P}\left[\left\{\text{Swap is accepted}\right\}\cap 
\left\{\widetilde{Y}'=\tilde{y}_l\right\} \cap \left\{J=j\right\}\right]
\end{multline*}
is symmetric.
\begin{multline*}
R_j\left((x_l,x_{l+1}),(y_l,y_{l+1})\right)
= \pi_l\left(x_l\right)\pi_{l+1}\left(x_{l+1}\right)\\
 \times\int
 \min\biggl\lbrace 1,\ \frac{ \pi_{l+1}(\hat{x}_l)\sum_{k=1}^{M} 
W^k_U}
{\pi_{l+1}(x_{l+1})
\sum_{k=1}^{M} W^k_V}\biggr\rbrace\frac{W_U^k}{\sum_{k=1}^{M}W^k_U}\\
\times
p_l^{M}\left((v^{j+1},\dots,v^{M})\vert (\tilde{y}_l,v^1,\dots,v^j),\ \hat{x}_l\right)
p_l^{M}\left((u^1,\dots,u^{M})\vert \tilde{x}_l,\ x_{l+1}\right)\\
\times\delta(u^j-\tilde{y}_l)\delta(v^j-\tilde{x}_l)
\left(\prod_{1\leq k<j} \delta(u^{j-k}-v^k) \right)\left(\prod_{ {k>1},\ {k\neq j} }du^k dv^k\right)
\end{multline*}
Recall the definition of the weights and the fact that $U^J=\tilde{y}_l,$
$V^J=\tilde{x}_l,$ 
and $V^j=U^{J-j}$ for $j=1,\dots,J-1$
\begin{align*}
W_U^j &= \pi_l\left(\tilde{y}_l,x_{l+1}\right) 
p_l^j\left(\left(u^{j-1},\dots,u^1,\tilde{x}_l\right)\vert \tilde{y}_l,\ \hat{x}_l\right)\\
&\hspace{24pt}\times\lambda^j\left(\left(\tilde{x}_l,u^1,\dots,u^{j-1},\tilde{y}_l\right),\  \hat{x}_l,x_{l+1}\right)\\
&= \pi_l\left(\tilde{y}_l,x_{l+1}\right) p_l^j\left(\left(v^1,\dots,v^{j-1}\right)\vert \tilde{y}_l,\ \hat{x}_l\right)\\
&\hspace{24pt}\times\lambda^j\left(\left(\tilde{x}_l,u^1,\dots,u^{j-1},\tilde{y}_l\right),\  \hat{x}_l,x_{l+1}\right).
\end{align*}
Thus,
\begin{multline*}
R_j\left((x_l,x_{l+1}),(y_l,y_{l+1})\right)
=  \pi_l\left(x_l\right)\pi_{l+1}\left(x_{l+1}\right)
\\\times \int
 \min\biggl\lbrace 1,\ \frac{ \pi_{l+1}(\hat{x}_l)\sum_{k=1}^{M} 
W^k_U}
{\pi_{l+1}(x_{l+1})
\sum_{k=1}^{M} W^k_V}\biggr\rbrace
\frac{1}{\sum_{k=1}^{M}W^k_U}\ \pi_l\left(\tilde{y}_l,x_{l+1}\right) \\
\times
p_l^j\left(\left(v^1,\dots,v^{j-1}\right)\vert \tilde{y}_l,\ \hat{x}_l\right)
\lambda^j\left(\left(\tilde{x}_l,u^1,\dots,u^{j-1},\tilde{y}_l\right),\  \hat{x}_l,x_{l+1}\right)\\
\times p_l^{M}\left((v^{j+1},\dots,v^{M})\vert (\tilde{y}_l,v^1,\dots,v^j),\ \hat{x}_l\right)
p_l^{M}\left((u^1,\dots,u^{M})\vert \tilde{x}_l,\ x_{l+1}\right)\\
\times\delta(u^j-\tilde{y}_l)\delta(v^j-\tilde{x}_l)
\left(\prod_{1\leq k<j} \delta(u^{j-k}-v^k) \right)\left(\prod_{ {k>1},\ {k\neq j} }du^k dv^k\right)
\end{multline*}
Definition \eqref{prodp} implies that for all $j$,
\begin{multline*}
 p_l^j\left(\left(v^1,\dots,v^{j-1}\right)\vert \tilde{y}_l,\ \hat{x}_l\right)
 p_l^{M}\left((v^{j+1},\dots,v^{M})\vert (\tilde{y}_l,v^1,\dots,v^{j-1},\tilde{x}_l),\ \hat{x}_l\right)\\ =
 p_l^{M}\left(\left(v^1,\dots,v^{M}\right)\vert \tilde{y}_l,\ \hat{x}_l\right),
 \end{multline*}
 Thus,
 \begin{multline*}
R_j\left((x_l,x_{l+1}),(y_l,y_{l+1})\right)
=\\ \pi_l\left(x_l\right)\pi_{l+1}\left(x_{l+1}\right)
 \int
 \min\biggl\lbrace 1,\ \frac{ \pi_{l+1}(\hat{x}_l)\sum_{k=1}^{M} 
W^k_U}
{\pi_{l+1}(x_{l+1})
\sum_{k=1}^{M} W^k_V}\biggr\rbrace\\
\times\frac{\pi_l\left(\tilde{y}_l,x_{l+1}\right)
\lambda^j\left(\left(\tilde{x}_l,u^1,\dots,u^{j-1},\tilde{y}_l\right),\  \hat{x}_l,x_{l+1}\right)}
{\sum_{k=1}^{M}W^k_U}\\
\times p_l^{M}\left((v^1,\dots,v^{M})\vert \tilde{y}_l,\ \hat{x}_l\right)
p_l^{M}\left((u^1,\dots,u^M)\vert \tilde{x}_l,\ x_{l+1}\right)\\
\times\delta(u^j-\tilde{y}_l)\delta(v^j-\tilde{x}_l)
\left(\prod_{1\leq k<j} \delta(u^{j-k}-v^k) \right)\left(\prod_{ {k>1},\ {k\neq j} }du^k dv^k\right)
\end{multline*}
which can be rewritten,
\begin{multline*}
R_j\left((x_l,x_{l+1}),(y_l,y_{l+1})\right)
=\pi_l\left(x_l\right)\pi_{l+1}\left(x_{l+1}\right)\pi_l\left(\tilde{y}_l,x_{l+1}\right)
\pi_{l+1}\left(\hat{x}_l\right)\\
\times\int
 \min\biggl\lbrace \frac{1}
{\pi_{l+1}(\hat{x}_l)
\sum_{k=1}^{M} W^k_U},\ \frac{1}
{\pi_{l+1}(x_{l+1})
\sum_{k=1}^{M} W^k_V}\biggr\rbrace
\\
\times\lambda^j\left(\left(\tilde{x}_l,u^1,\dots,u^{j-1},\tilde{y}_l\right),
\  \hat{x}_l,x_{l+1}\right)\\
\times p_l^{M}\left((v^1,\dots,v^{M})\vert \tilde{y}_l,\ \hat{x}_l\right)
p_l^{M}\left((u^1,\dots,u^{M})\vert \tilde{x}_l,\ x_{l+1}\right)\\
\times
\delta(u^j-\tilde{y}_l)\delta(v^j-\tilde{x}_l)
\left(\prod_{1\leq k<j} \delta(u^{j-k}-v^k) \right)\left(\prod_{ {k>1},\ {k\neq j} }du^k dv^k\right)
\end{multline*}
Plugging $y_{l+1} = \hat{x}_l$ and $\hat{y}_l = x_{l+1},$ into this expression
yields,
\begin{multline*}
R_j\left((x_l,x_{l+1}),(y_l,y_{l+1})\right) =
\pi_l\left(x_l\right)\pi_{l+1}\left(x_{l+1}\right)\pi_l\left(y_l\right)
\pi_{l+1}\left(y_{l+1}\right)\\
\times\int
 \min\biggl\lbrace \frac{1}
{\pi_{l+1}(y_{l+1})
\sum_{k=1}^{M} W^k_U},\ \frac{1}
{\pi_{l+1}(x_{l+1})
\sum_{k=1}^{M} W^k_V}\biggr\rbrace
\\
\times\lambda^j\left(\left(\tilde{x}_l,u^1,\dots,u^{j-1},\tilde{y}_l\right),
\  y_{l+1}, x_{l+1}\right)\\
\times p_l^{M}\left((v^1,\dots,v^{M})\vert \tilde{y}_l,\ y_{l+1}\right)
p_l^{M}\left((u^1,\dots,u^{M})\vert \tilde{x}_l,\ x_{l+1}\right)\\
\times
\delta(u^j-\tilde{y}_l)\delta(v^j-\tilde{x}_l)
\left(\prod_{1\leq k<j} \delta(u^{j-k}-v^k) \right)\left(\prod_{ {k>1},\ {k\neq j} }du^k dv^k\right)
\end{multline*}
By the symmetry property of $\lambda_j$  this expression is
symmetric in $\left(x_l,x_{l+1}\right)$ and $\left(y_l,y_{l+1}\right)$.
\end{proof}

Clearly a Markov chain that evolves only by swap moves cannot
sample all configurations, ie. the chain generated by $\psi$ is
not $\phi$-irreducible for any non trivial measure $\phi.$
These swap moves must therefore  
be used in conjunction with a transition rule that can
reach any region of space.  More precisely, let
 $\tau$ from expression \eqref{def:tau} be Harris recurrent with stationary distribution $\Pi$
 (see \cite{meyn93}).
The the transition rule for parallel marginalization is
\begin{equation*}
\tau_{pm}(y\vert x) = (1-\alpha)\ \tau(y\vert x)\\
+ \alpha\ \int \tau(z\vert x)\psi\left(y\vert z\right)dz
\end{equation*}
where
$$
\psi(y\vert x) = \sum_{k=0}^{L-1} \frac{1}{L} \psi^M_l\left(y\vert x\right)
$$
and $\alpha\in \left[0,1\right)$ is the probability that a swap move occurs.
$\tau_{pm}$ dictates that, with probability $\alpha$, the chain
attempts a swap move between levels $I$ and $I+1$ where $I$ is a random
variable chosen uniformly from $\left\{0,\dots,L-1\right\}$.
Next, the chain evolves according to  $\tau$.  
With probability
$1-\alpha$ the chain moves only according to $\tau$ and does not attempt a swap.
The next result guarantees the
invariance of $\Pi$ under evolution by $\tau_{pm}$.


It is not difficult to verify that the chain generated by $\tau_{pm}$ has invariant 
measure $\Pi$ and is Harris recurrent if the chain generated by $\tau$ has these properties.
Thus by combining standard MCMC steps on each component,
governed by the transition probability $\tau$, with swap steps between
the components governed by $\psi$, an MCMC method results
that not only uses information from rapidly equilibrating
lower dimensional chains, but is also convergent.

\section{Numerical Examples}
In this section I consider applications of parallel marginalization to two conditional path sampling problems for a one dimensional stochastic differential equation,
 \begin{equation}\label{sde1}
dZ(t) = f\left(Z(t)\right)dt
+ \sigma\left(Z(t)\right)dW(t),
\end{equation}
where $f$ and $\sigma$ are real valued functions of $\mathbb{R}.$
One must
first approximate $Z(t)$ by a discrete process 
for which the path density is readily available.
Let $t_0=0,t_1=\frac{T}{N},\dots,t_N=T$ be a mesh on which one wishes to calculate path
averages.
One such approximate process is given by the 
linearly implicit
Euler scheme (a balanced implicit method, see \cite{milstein98}),
\begin{equation}\label{def:lie}
\begin{split}
&X(n+1) = X(n) + f\left(X(n)\right)\triangle\\
&\ \ \ \ + \left(X(n+1) - X(n)\right)f^{'}\left(X(n)\right)\triangle
+ \sigma\left(X(n)\right)\sqrt{\triangle}\ \xi(n),\\
&X(0) = Z(0).
\end{split}
\end{equation}
Here $X(n)$ is an approximation to $Z$ at time $t_n.$
The reader should note that the rate of convergence of the above scheme to 
the solution of \eqref{sde1} would not be effected by the
insertion  in \eqref{def:lie} of a non-negative constant in front of the $f^{'}$ term.  The choice
of $1$ made here seemed to improve the stability of the resulting scheme for large values
of $\triangle.$
The $\xi(n)$ are independent Gaussian random
variables with mean 0 and variance 1,
and $\triangle=\frac{T}{N}.$  $N$ is assumed to be a power
of 2.
The choice of this scheme over the Euler scheme (see \cite{kloeden92}) 
is due to its favorable stability properties as explained later.  It is henceforth assumed that
$X(t)$ instead of $Z(t)$ is the process of interest.

The first of the conditional sampling problems discussed here 
is the bridge sampling problem in which one 
generates samples  of transition paths between two states.
This problem arises, for example, in financial volatility estimation where, given a sequence
of observations, $\left(z(s_0),\dots,z(s_K)\right)$ with $\left\{s_j\right\}\subset \left\{t_l\right\},$
the goal is to estimate the diffusion
term $\sigma$ (assumed here to be constant) appearing in the stochastic differential
equation.  Since in general one cannot easily evaluate the transition probability between
times $s_j$ and $s_{j+1}$ (and thus the likelihood of the observations) 
it is necessary to generate samples between the observations,
$$
V(j) = \left(X(j,1),\dots,X(j,N_j)\right)
$$
where $N_j= N\left(s_{j+1}-s_j\right) -1$ (assumed to be an integer) 
and $X(j,n)$ denotes the value of the process at time $s_j+\frac{n}{N_j}$.  
It is then easy to evaluate the likelihood
of a path
$$
X(s_0),V(0),\dots,X(s_K),V(K)
$$
given a particular value of the volatility, $\sigma.$

The filtering/smoothing problem is similar to the financial volatility example of the previous
paragraph except that now it is  assumed that the observations are noisy functions of the
underlying process.  For example, one may wish to sample possible trajectories taken by a rocket
given somewhat unreliable GPS observations of its position.  If the conditional density of 
the observations given the position of the rocket is known,
 it is possible to generate conditional samples
of the trajectories.

\subsection{Bridge path sampling}
In the bridge path sampling problem
one seeks to approximate conditional expectations of the form
$$
\mathbf{E}\left[g\left(Z(t_1),\dots,Z(t_{N-1})\right)\ \vert
\{{Z(0)=z^-}\},\{{Z(T)=z^+}\}\right]
$$
where $g$ is a real valued function, and
 $Z(t)$ is solution to 
\eqref{sde1}.

Without the condition $Z(T)=z^+$ above, generating an approximate sample
$\left(X(0),\dots,X(N)\right)$ path is a relatively straitforward
endeavor.
One simply generates a sample of $Z(0)$,
then evolves \eqref{def:lie} with this initial condition.  
However, the presence of
information about $\left\{Z(t)\right\}_{t>0}$ complicates 
the task.
In general, some sampling method which requires only 
knowlege of a function
proportional to conditional density of  
$\left(X(1),\dots,X(N-1)\right)$
must be applied.
The approximate path density associated with discretization
\eqref{def:lie} is
\begin{align}\label{def:pi_0:ex1}
\pi_0\left(x_0(1),\dots,x_0(N-1)\ \vert x_0(0),x_0(N)\right)&\\
 &\hspace{-70pt}\propto
\exp\left(-\sum_{k=0}^{N-1} 
\mathcal{V}\left(x_0(n),x_0(n+1),\triangle\right)
\right)\notag
\end{align}
where
\begin{equation*}\label{def:V}
\mathcal{V}\left(x,y,\triangle\right) = 
\frac{\left[\left(1-\triangle f^{'}\left(x\right)\right)\left(y
-x\right)+\triangle f\left(x\right)\right]^2}
{2\sigma^2\left(x\right)\triangle}
\end{equation*}

At this point the parallel marginalization sampling
procedure is applied to the density $\pi_0$.
However, as indicated above, a prerequisite for the use of parallel 
marginalization is the ability to estimate marginal densities.
In some important problems
 homogeneities in the underlying system yield 
simplifications in the calculation of these densities by the
methods in \cite{chorin03,stinis05}.
These calculations
can be carried out before implementation of parallel marginalization,
or they can be integrated into the sampling procedure.

In some cases, computer generation of the $\left\{\pi_l\right\}_{l>0}$
can be completely avoided.  
The examples presented here are two such cases.
Let \linebreak $S_l = \left\{0,2^l,2(2^l),3(2^l),\dots,N\right\}$ (recall $N$ is a power of 2).  
Decompose $S_l$
as $\widehat{S}_l\sqcup\widetilde{S}_l$ where
$$\widehat{S}_l = \left\{0,2(2^l),4(2^l),6(2^l),\dots,N\right\}$$
 and
$$\widetilde{S}_l = \left\{2^l,3(2^l),5(2^l),7(2^l),\dots,N-2^l\right\}.$$
In the notation of the previous sections, $x_l=\left(\hat{x}_l,\tilde{x}_l\right)$
where \linebreak $\hat{x}_l=\left\{x_l(n)\right\}_{n\in\widehat{S}_l\setminus\left\{0,N\right\}}$ and
$\tilde{x}_l=\left\{x_l(n)\right\}_{n\in\widetilde{S}_l}.$
In words, the hat and tilde variables represent alternating time slices of the path.
For all $l$ fix $x_l(0)=z^-$ and $x_l(N)=z^+$.
We choose the approximate marginal densities
$$
\pi_l\left(\left\{x_l(n)\right\}_{n\in S_l\setminus\left\{0,N\right\}}\ 
\vert x_l(0),x_l(N)\right) 
\propto q_l\left(\left\{x_l(n)\right\}_{n\in S_l}\right)
$$ 
where
for each $l$, $q_l$ is defined by
successive coarsenings of \eqref{def:lie}.
That is,
\begin{equation*}\label{def:q_l}
q_l\left( \left\{x_l(n)\right\}_{n\in S_l} \right)
= \exp\left(-\sum_{k=0}^{N/2^l-1} 
\mathcal{V}\left(x_l(2^lk),x_l(2^l(k+1)),2^l\triangle\right)\right).
\end{equation*} 
Since $\pi_l$ will be sampled
using a Metropolis-Hastings method 
with  $x(0)$ and $x(N)$ fixed,
knowlege of the normalization constants
$$
\mathcal{Z}_l\left(x_l(0),x_l(N)\right) 
= \int q_l\ \prod_{n\in S_l\setminus\left\{0,N\right\}}dx_l(n)
$$
is unnecessary.

Notice from \eqref{def:pi_0:ex1} that, conditioned on the values of $X(n-1)$ and
$X(n+1)$, the variance of $X(n)$ is of order $\triangle$.  Thus
any perturbation of $X(n)$ which leaves $X(m)$ fixed for $m\neq n$ and
which is compatible with joint distribution \eqref{def:pi_0:ex1} must be of the
order $\sqrt{\triangle}$.  This suggests that distributions defined by
coarser discretizations of \eqref{def:pi_0:ex1} will allow larger perturbations,
and consequently will be easier to sample.  However, it is important to 
choose a discretization that remains stable for large values of $\triangle$.
For example, while the linearly implicit Euler method performs well in 
the experiments below, similar tests
using the Euler method were less
successful due to limitations on the largest allowable values of $\triangle$.

In this numerical example 
bridge paths are sampled between time 0 and time 10 for a diffusion in 
a double well potential
$$
f(x) = -4x\left(x^2-1\right)\ \ \text{and}\ \ \sigma(x) = 1
$$
The left and right end points are chosen
as $z^-=z^+=0$.  $\triangle = 2^{-10}.$
$Y^n_l\in\mathbb{R}^{10/\left(2^l\triangle\right)+1}$
is the $l^{th}$ level of the parallel marginalization 
Markov chain at algorithmic time $n$.
There are 10 chains ($L = 9$).
The observed swap acceptance rates 
are reported in table \eqref{swaprates1}.  Notice that the swap rates are highest
at the lower levels but seems to stabilize at the higher levels.

\begin{threeparttable}
\caption{Swap acceptance rates for bridge sampling
problem}
\begin{tabular*}{\hsize}{@{\extracolsep{\fill}}rrrrrrrrrrrrr}
\hline
\multicolumn1c{Levels\tnote{1}}&
\multicolumn1c{0/1}&\multicolumn1c{1/2}&
\multicolumn1c{2/3}&\multicolumn1c{3/4}&
\multicolumn1c{4/5}&\multicolumn1c{5/6}&
\multicolumn1c{6/7}&\multicolumn1c{7/8}
&\multicolumn1c{8/9}
\cr\hline
\multicolumn1c{}&
\multicolumn1c{0.86}&\multicolumn1c{0.83}&
\multicolumn1c{0.75}&\multicolumn1c{0.69}&
\multicolumn1c{0.54}&\multicolumn1c{0.45}&
\multicolumn1c{0.30}&\multicolumn1c{0.22}
&\multicolumn1c{0.26}
\cr
\hline
\end{tabular*}
\begin{tablenotes}
\item[1] Swaps between levels $l$ and $l+1.$
\end{tablenotes}
\label{swaprates1}
\end{threeparttable}
\vspace{1cm}

\begin{figure}[ht] \centering
\includegraphics[width=4in]{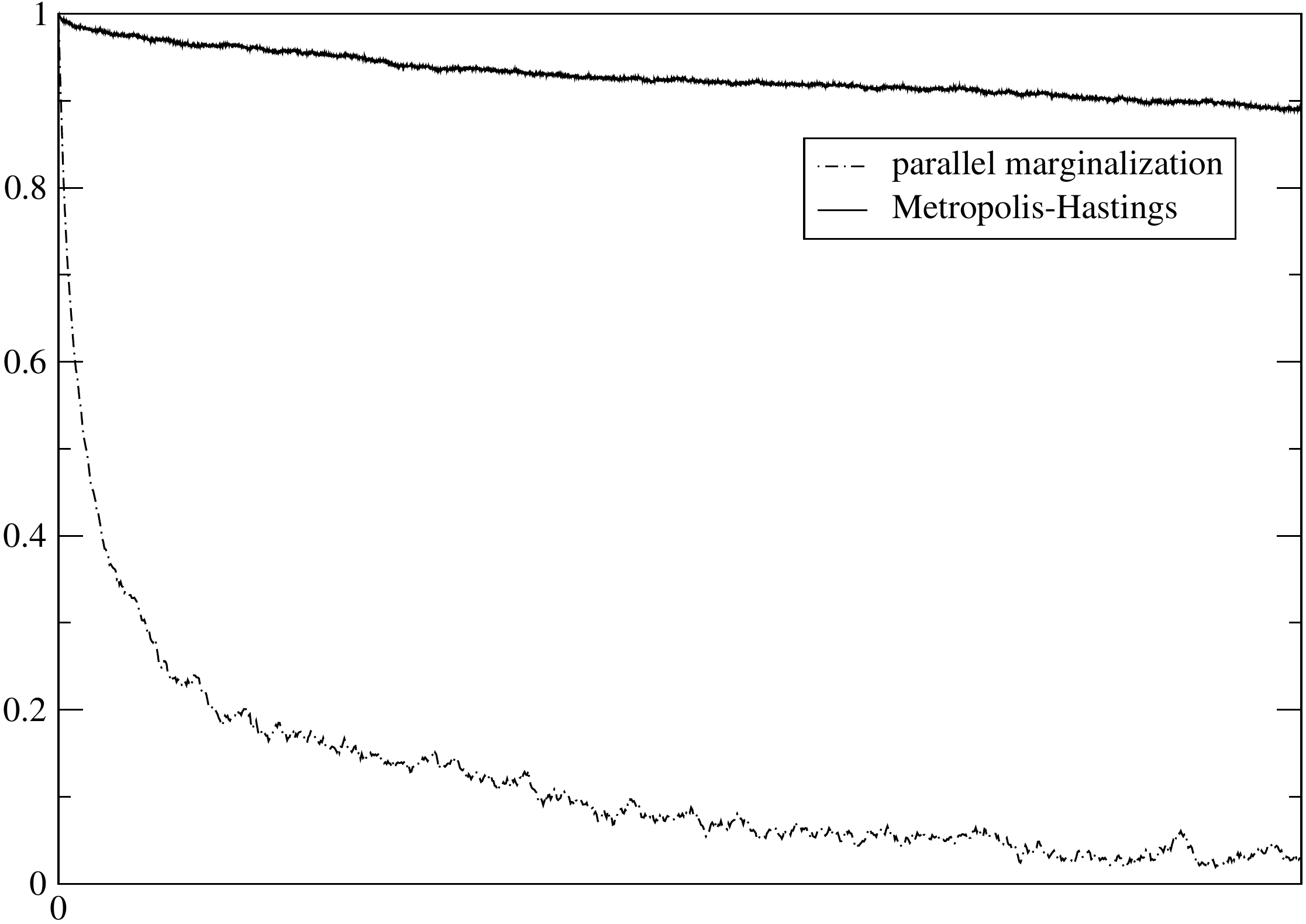}
\caption{  Autocorrelation of $Y^n_{mid}$ for Metropolis-Hastings method with
1-d Gaussian random walk proposals (solid) and parallel marginalization (dotted).
The x-axis runs from 0 to 10000 iterations of the Metropolis-Hastings method and
from 0 to 1000 iterations of parallel marginalization.  This rescaling more than
compensates for the extra work for parallel marginalization per iteration. }
\label{fig1}
\end{figure}

Let $Y^n_{mid}\in\mathbb{R}$ denote the midpoint of the path defined by $Y^n_0$
(i.e. an approximate sample of the path at time 5).  
In Figure \ref{fig1} the autocorrelation of $Y^n_{mid}$
$$
\mathbf{Corr}\left[Y^n_{mid},Y^0_{mid}\right]
$$
is compared to that of a standard Metropolis-Hastings rule using 1 dimensional 
Gaussian random walk proposals.
In the figure, the time scale of the autocorrelation for the Metropolis-Hastings method has
been scaled by a factor of 1/10 to more than account for the extra computational time
required per iteration of parallel marginalization.
The relaxation time of the parallel chain is clearly reduced.

In these numerical examples,  parallel marginalization is applied
with a slight simplification as detailed in the following algorithm. 
\begin{alg}
The chain moves from $Y^n$ to $Y^{n+1}$ as follows:
\begin{enumerate}
\item 
 Generate M independent Gaussian random paths 
$\left\{\zeta^m\left(n\right)\right\}_{n\in\widetilde{S}_l}$  with independent components
$\zeta^m\left(n\right)$ of mean 0 and variance $2^{l-1}\triangle$.
\item
For each $j$ and $n\in\widetilde{S}_l$ let 
$$
U^m\left(n\right) = \zeta^m\left(n\right)+0.5\left(x_{l+1}(n-1)+x_{l+1}(n+1)\right).
$$
\item Define the weights
$$
W_U^m = \frac{\pi_l\left(x_{l+1},U^m\right)}
{p_l\left(U^m\vert x_{l+1}\right)},
$$
where $p_l$ is defined by the choice in step 1 as
$$
p_l\left(\tilde{x}_l\vert \hat{x}_l\right)
\propto
\exp\left(
\sum_{n\in\widetilde{S}_l} 
-\frac{\left(
\tilde{x}_l(n)
-0.5\left(\hat{x}_l(n-1)+\hat{x}_l(n+1)\right)
\right)^2}
{2^l\triangle}
\right).
$$
\item Choose $J\in\left\{1,\dots,M\right\}$ according to the
probabilities
$$
\mathbf{P}\left[ J = j \right] 
= \frac{W_U^j}{\sum_{k=1}^{M} W_U^k}.
$$
Set $\widetilde{Y}' = U^J.$
\item Set $V^J = \tilde{x}_l$ and for $j\neq J$ set
$$V^j\left(n\right)=
\zeta^j\left(n\right)+0.5\left(\hat{x}_l(n-1)+\hat{x}_l(n+1)\right).
$$
\item Define the weights
$$
W_V^m = \frac{\pi_l\left(\hat{x}_l,V^m\right)}
{p_l\left(V^m\vert\hat{x}_l\right)}.
$$

\item
Set $$Y^{n+1} = \left(\dots,x_{l+1},\tilde{y}_l,\hat{x}_l,\dots\right)$$
with probability
\begin{equation}\label{def:A^M_l}
A^M_l = \min\biggl\lbrace 1,\ \frac{ \pi_{l+1}(\hat{x}_l)\sum_{m=1}^{M} 
W^m_U}
{\pi_{l+1}(x_{l+1})
\sum_{m=1}^{M} W^m_V}\biggr\rbrace
\end{equation}
and
$$Y^{n+1} = Y^n=\left(\dots,\hat{x}_l,\tilde{x}_l,x_{l+1},\dots\right)$$
with probability
$
1-A^M_l
$.
\end{enumerate}
\end{alg}
This simplification reduces by half the number of gaussian random variables needed
to evaluate the acceptance probability but may not be appropriate in all settings.
For this
problem, the choice of $M$ in \eqref{def:A^M_l}, 
the number of samples $U^m$ 
and $V^m$, seems to have little effect on the
swap acceptance rates.  In the numerical experiment $M=l+1$
for swaps between levels $l$ and $l+1$.

\begin{figure}[ht] \centering
\includegraphics[width=4in]{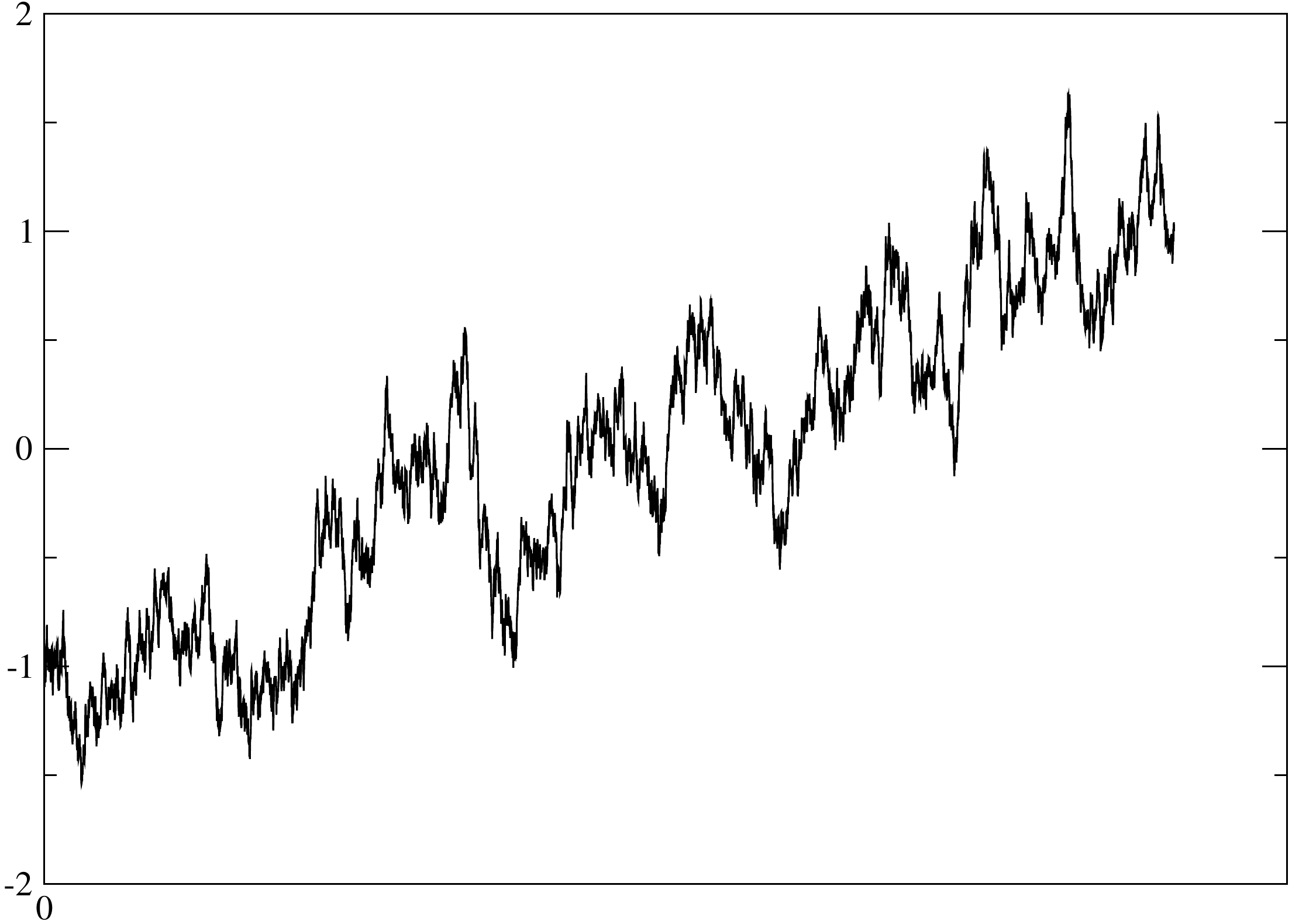}
\caption{   Metropolis generated bridge path from Section 3.1 after a 10 minute run on a standard
desktop workstation.  Clearly the method has not converged.}
\label{bridgesample}
\end{figure}

\begin{figure}[ht] \centering
\includegraphics[width=4in]{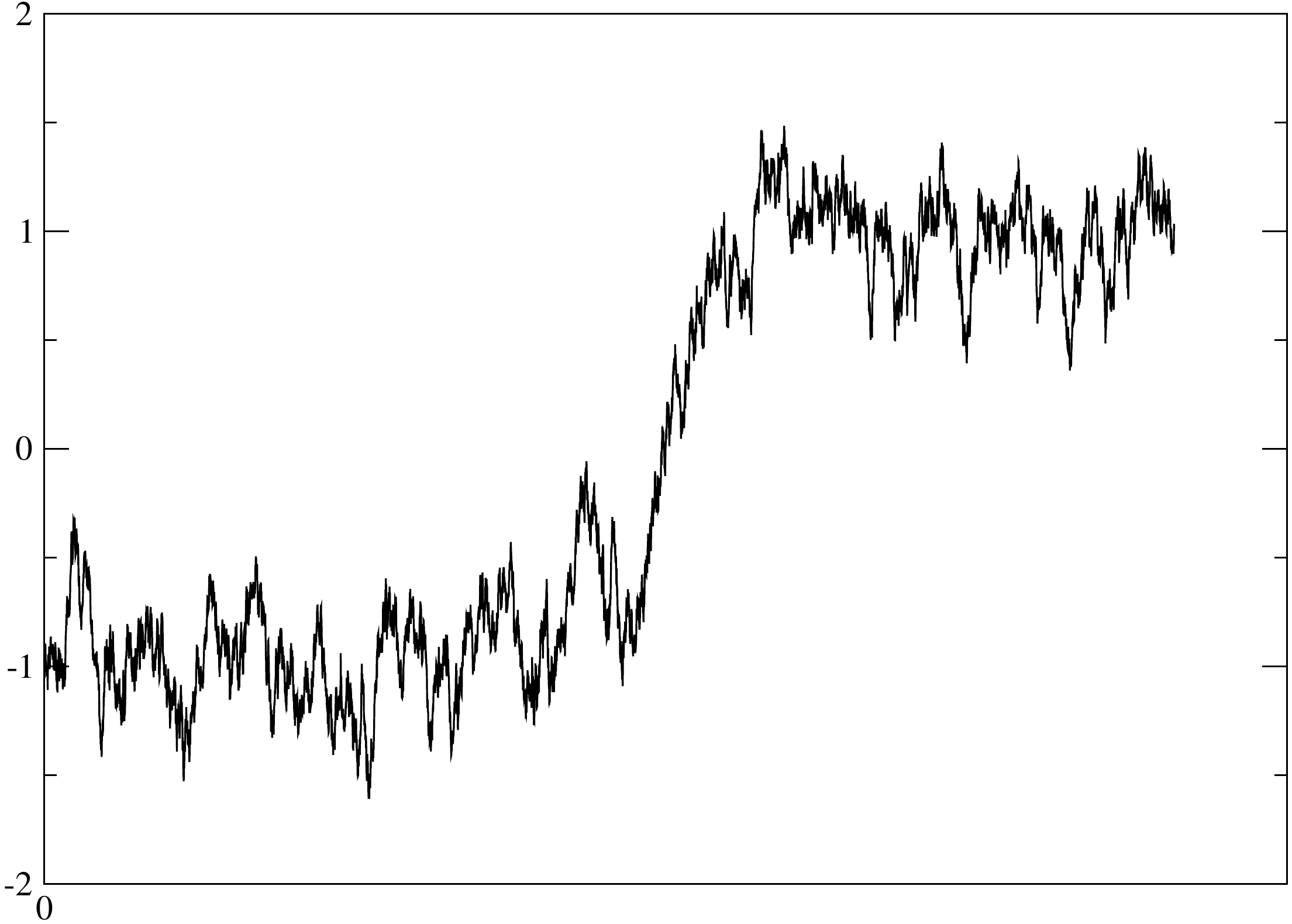}
\caption{  Parallel marginalization generated bridge path from Section 3.1 after a 10 minute run on a standard
desktop workstation.  Apparently the method has converged.}
\label{parbridge}
\end{figure}

The results of the Metropols-Hastings and parallel marginalization methods applied to the
 above bridge sampling problem after a run time of 10 minutes on a standard workstation
are presented in Figures \ref{bridgesample} and \ref{parbridge}.  Apparently the
sample generated by parallel marginalization is a reasonable bridgepath while
the Metropolis-Hastings method has clearly not converged.

\subsection{Non-linear smoothing/filtering}
In the non-linear smoothing and filtering problem
one seeks to approximate conditional expectations of the form
$$
\mathbf{E}\left[g\left(Z(0),Z(t_1),\dots,Z(T)\right)  \ \vert
\left\{H(j)=h(j)\right\}_0^K\right]
$$
where
the real valued processes
$Z(t)$ and $H(j)$ are given by
the system
\begin{equation}\label{filterproblem}
\begin{split}
&dZ(t) = f\left(Z(t)\right)dt
+ \sigma\left(Z(t)\right)dW(t), \\
&H(j) = r\left(Z(s_j)\right) + \chi(j),\\
&Z(0) \sim \rho,\ \ \ \ \chi(j)\sim i.i.d.\ \mu.
\end{split}
\end{equation}
$g$, $f$, $\sigma$, and $r$ are real valued functions of $\mathbb{R}$.
The $\chi(j)$ are
real valued independent random variable drawn from the density $\mu$ and are
independent of the Brownian motion $W(t).$
$\left\{s_j\right\}\subset\left\{t_j\right\},$ 
and $0=s_0<s_1<...<s_K=T.$
The process $Z(t)$ is a hidden signal and the $H(j)$ are 
noisy observations.  The idea of computing the above conditional expectation by
conditional path sampling has been suggested in \cite{alexander04,stuart04}.  Popular alternatives include
particle filters (see \cite{defreitas05}) and ensemble Kalman filters (see \cite{evensen03}).

Again, begin by discretizing the system.  Assume that
$N_j= N\left(s_{j+1}-s_j\right) -1$ is an integer and let $\triangle=\frac{T}{N}.$
The
linearly implicit Euler scheme gives
\begin{equation*}
\begin{split}
&X(j,n+1) = X(j,n) + f\left(X(j,n)\right)\triangle\\
&\ \ \ \ + \left(X(j,n+1) - X(j,n)\right)f^{'}\left(X(j,n)\right)\triangle
+ \sigma\left(X(j,n)\right)\sqrt{\triangle}\ \xi(j,n),\\\
&H(j) = r\left(X(j)\right)+\chi(j),\notag\\
&X(0) = Z(0)\ \ \ \ \chi(j)\sim i.i.d.\ \mu
\end{split}
\end{equation*}
where $X(j,n)$ represents the discrete time approximation to $Z(s_j+n\triangle),$
for $0\leq n\leq N_j.$
The $\xi(n)$ are independent Gaussian random
variables with mean 0 and variance 1.
  The $\xi(n)$ are independent of the
$\chi^m$. $N$ is again assumed to be a power of 2.

The approximate path measure for this problem is
\begin{multline*}
\pi_0 \left(x_0(0),\dots,x_0(N)\ \vert
h(0),\dots,h(K)\right)\\
\propto 
\exp\left(
-\sum_{k=0}^{N-1}
\mathcal{V}\left(x_0(n),x_0(n+1),\triangle\right)
\right)\\
\times\rho\left(x_0(0)\right)\prod_{n=0}^{K} 
\mu\left(x_0(j)-r\left(h(j)\right)\right)
\end{multline*}
The approximate marginals are chosen as
\begin{multline*}
\pi_l \left(\left\{x_l(n)\right\}_{n\in S_l}\ 
\vert h(0),\dots,h(K)\right)
\propto\\
q_l\left(\left\{x_l(n)\right\}_{n\in S_l}\right)
\rho\left(x_l(0)\right)\prod_{n=0}^{K} \mu\left(x_l(j)-r\left(h(j)\right)\right)
\end{multline*}
where $V$, $q_l$ and $S_l$ are as defined in the previous section.

In this example, samples of the smoothed path are generated
between time
time 0 and time 10 for the same diffusion in 
a double well potential.  The densities $\mu$ and $\rho$ are chosen 
as
$$
\mu = N(0,0.01)\ \ \text{and}\ \ 
\rho(x) \propto \exp\left(-\left(x^2-1\right)^2\right)
$$
The function $r$ in \eqref{filterproblem} is the identity function.
The observation times are $s_0 = 0,s_1=1,\dots,s_{10}=10$
with
$H(j)=-1$ for $j=0,\dots,5$ and $H(j)=1$ for $j=6,\dots,10$.  $\triangle = 2^{-10}.$
There are 8 chains ($L = 7$).
The observed swap acceptance rates 
are reported in table \eqref{swaprates2}.
Notice that the swap rates are again highest
at the lower levels but, for this problem, become unreasonably small at the highest level.

\begin{threeparttable}
\caption{Swap acceptance rates for filtering/smoothing
problems}
\begin{tabular*}{\hsize}{@{\extracolsep{\fill}}rrrrrrrrrrrrr}
\hline
\multicolumn1c{Levels\tnote{1}}&
\multicolumn1c{0/1}&\multicolumn1c{1/2}&
\multicolumn1c{2/3}&\multicolumn1c{3/4}&
\multicolumn1c{4/5}&\multicolumn1c{5/6}&
\multicolumn1c{6/7}
\cr\hline
\multicolumn1c{}&
\multicolumn1c{0.86}&\multicolumn1c{0.83}&
\multicolumn1c{0.74}&\multicolumn1c{0.65}&
\multicolumn1c{0.46}&\multicolumn1c{0.23}&
\multicolumn1c{0.04}
\cr
\hline
\end{tabular*}
\begin{tablenotes}
\item[1] Swaps between levels $l$ and $l+1.$
\end{tablenotes}
\label{swaprates2}
\end{threeparttable}
\vspace{1cm}

\begin{figure}[ht] \centering
\includegraphics[width=4in]{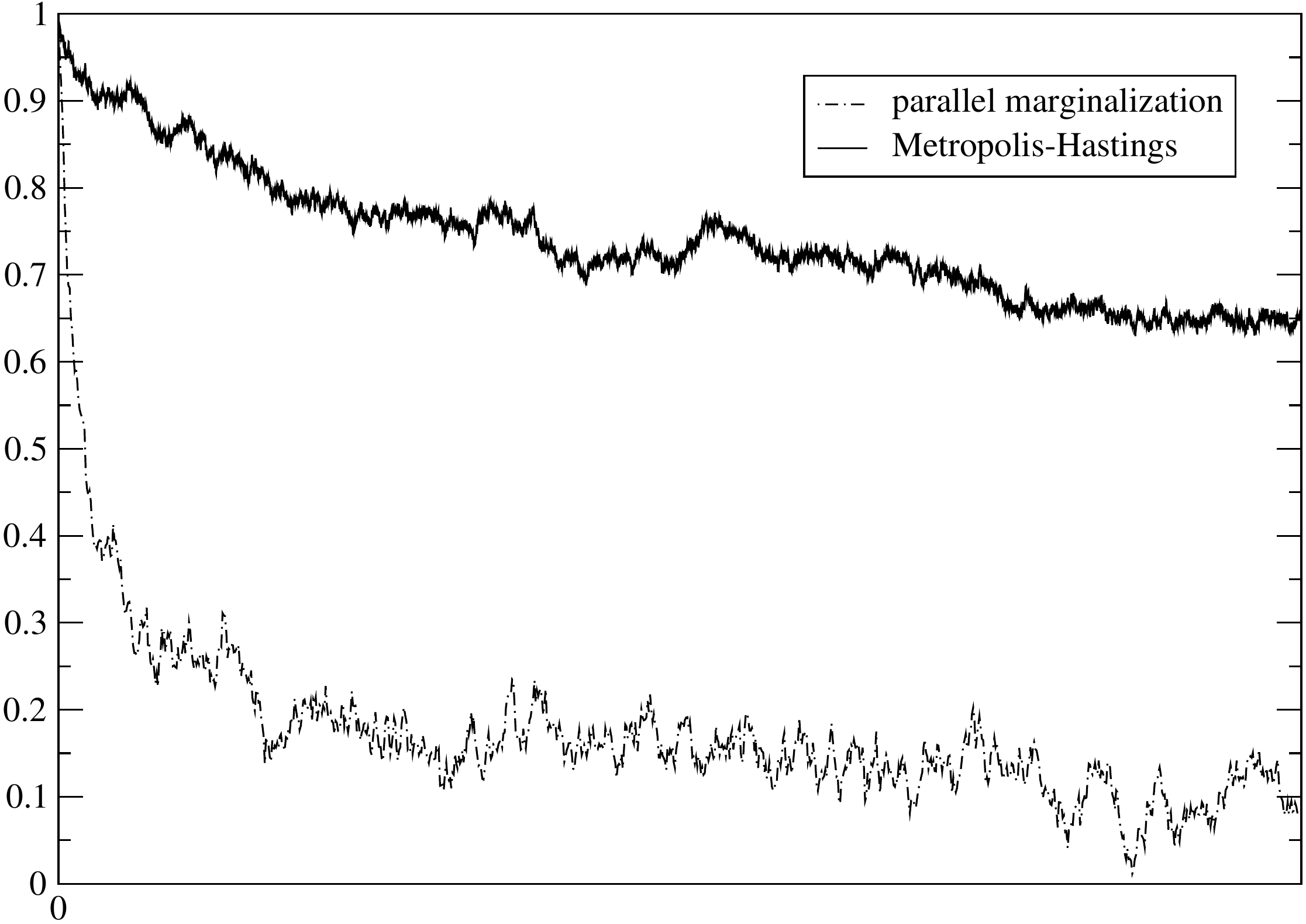}
\caption{  Autocorrelation of $Y^n_{mid}$ for Metropolis-Hastings method with
1-d Gaussian random walk proposals (solid) and parallel marginalization (dotted).
The x-axis runs from 0 to 10000 iterations of the Metropolis-Hastings method and
from 0 to 1000 iterations of parallel marginalization.  This rescaling more than
compensates for the extra work for parallel marginalization per iteration. }
\label{fig2}
\end{figure}

Again, $Y^n_{mid}\in\mathbb{R}$ denotes the
 midpoint of the path defined by $Y^n_0$
(i.e. an approximate sample of the path at time 5).  
In Figure \ref{fig2} the autocorrelation of $Y^n_{mid}$
is compared to that of a standard Metropolis-Hastings rule.
The figure has been adjusted as in the previous example.
The relaxation time of the parallel chain is again 
clearly reduced. 

The algorithm is modified as in the previous example. 
For this problem, acceptable 
swap rates require a higher choice of $M$ in \eqref{def:A^M_l} than needed
in the bridge sampling problem.  In this numerical experiment $M=2^l$
for swaps between levels $l$ and $l+1$. 

\begin{figure}[ht] \centering
\includegraphics[width=4in]{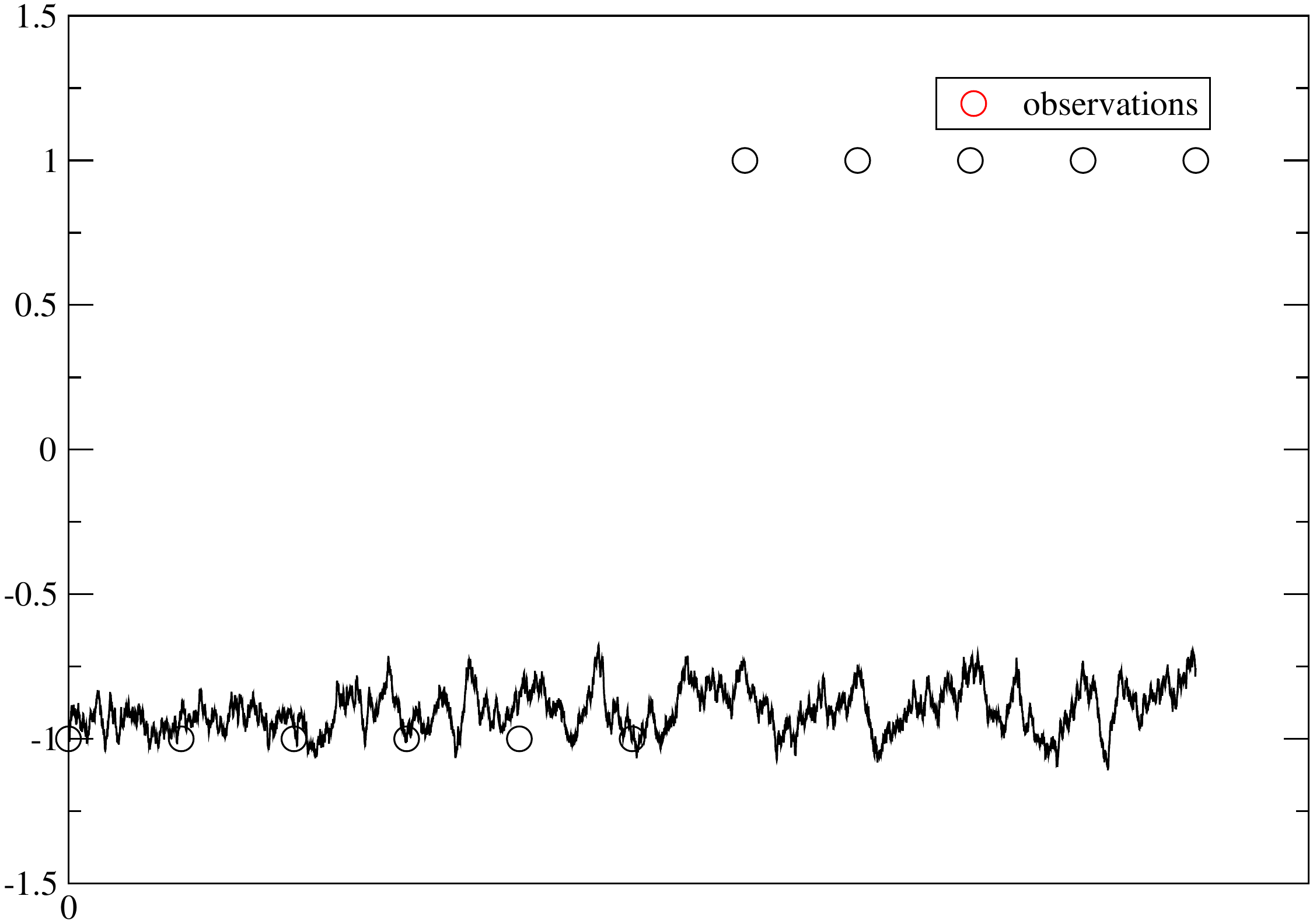}
\caption{   Metropolis-Hastings generated smoothed path from Section 3.2 after a 10 minute run on a standard
desktop workstation.  Clearly the method has not converged.}
\label{pathsmooth}
\end{figure}

\begin{figure}[ht] \centering
\includegraphics[width=4in]{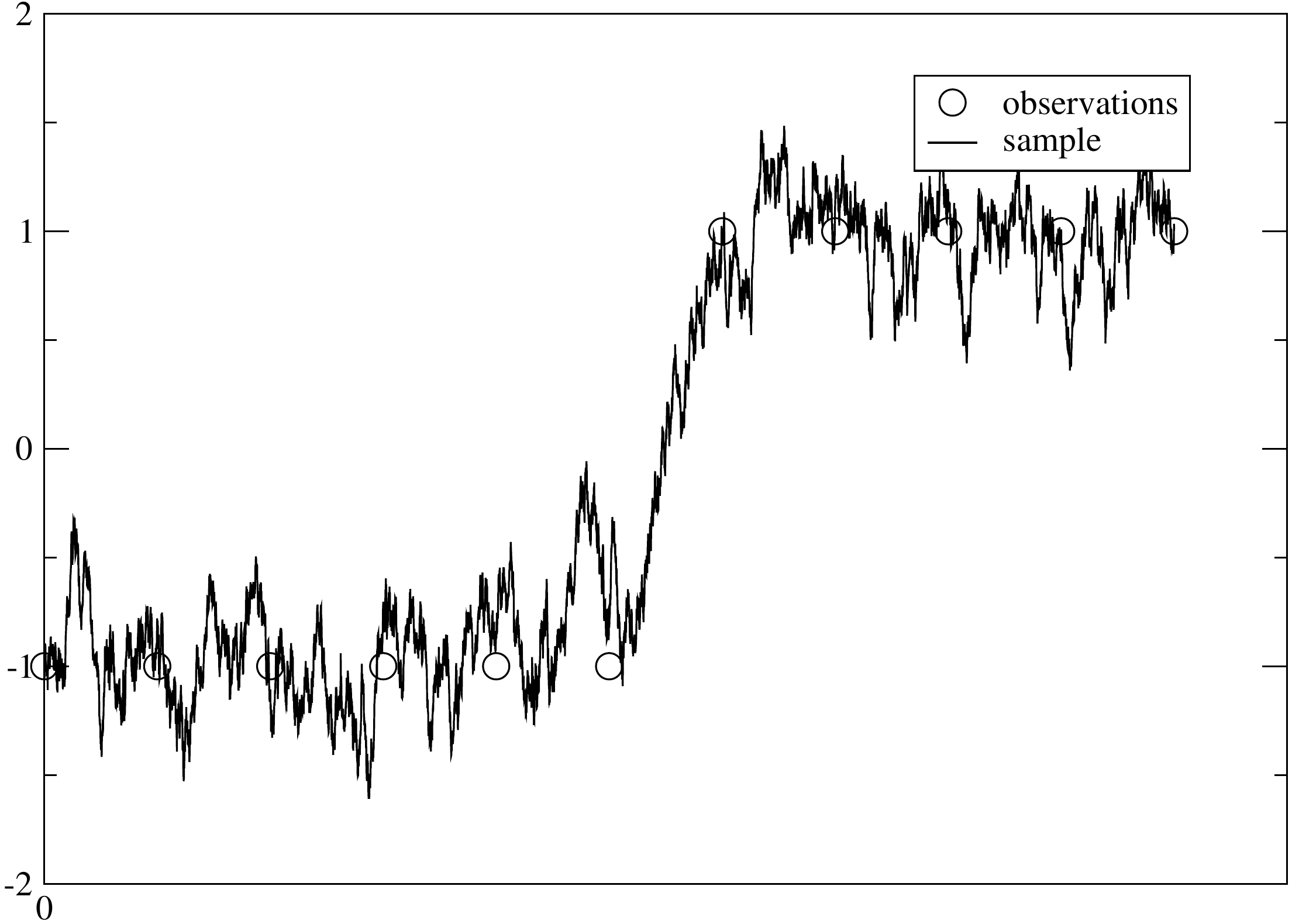}
\caption{  Parallel marginalization generated smoothed path from Section 3.2 after a 10 minute run on a standard
desktop workstation.  Apparently the method has converged.}
\label{parsmooth}
\end{figure}

The results of the Metropols-Hastings and parallel marginalization methods applied to the
smoothing problem  above after a run time of 10 minutes on a standard workstation
are presented in figure \ref{pathsmooth} and \ref{parsmooth}.  Apparently the
sample generated by parallel marginalization is a reasonable bridgepath while
the Metropolis-Hastings method has clearly not converged.

\section{Conclusion}
A Markov chain Monte Carlo method has been proposed and applied to two 
conditional path sampling problems for stochastic differential equations.
Numerical results indicate that this method, parallel marginalization,
can have a dramatically reduced equilibration time when compared to standard MCMC 
methods.  

Note that parallel marginalization should not be viewed as 
a stand alone method.  Other acceleration techniques such as hybrid Monte Carlo 
can and should be implemented at each level within
the parallel marginalization framework.
As the smoothing problem indicates, the acceptance probabilities at
coarser levels can become small.  The remedy for this is the development of 
more accurate approximate marginal distributions by, for example,
the methods
in \cite{chorin03} and \cite{stinis05}.

\section{Acknowledgments}
I would
like to thank Prof. A. Chorin for his guidance during this research, which
was carried out while I was a  Ph.D. student at U. C. Berkeley.
I would also like to thank Prof. O. Hald, Dr. P. Okunev, Dr. P. Stinis, and Dr. Xuemin Tu,
for their very helpful comments.
This work was supported by the Director, Office of Science, Office
of Advanced Scientific Computing Research, of the U. S. Department of
Energy under Contract No. DE-AC03-76SF00098 and National Science Foundation
grant DMS0410110.


\end{document}